\documentclass[a4paper,11pt]{article}
\pdfoutput=1 

\usepackage{jcappub}
\usepackage[T1]{fontenc}
\usepackage{amssymb,amsmath,amsfonts,amsbsy,graphicx,microtype,rotating,bm,todonotes}
\usepackage{hyperref}
\def\be{\begin{equation}}
\def\ee{\end{equation}}
\def\ba{\begin{eqnarray}}
\def\ea{\end{eqnarray}}
\renewcommand{\(}{\left(}
\renewcommand{\)}{\right)}
\renewcommand{\[}{\left[}
\renewcommand{\]}{\right]}

\newcommand{\mpl}{M_{\rm pl}}
\newcommand{\fnl}{f_{\rm NL}}

\newcommand{\calR}{\mathcal{R}}

\usepackage{color}
\newcommand{\tc}{\textcolor{blue}}

\def\nn{\nonumber}

\rightline{MIT-CTP/4974, YITP-17-133}

\title{\boldmath Revisiting non-Gaussianity from non-attractor inflation models}

\author[a]{Yi-Fu Cai,}
\author[b]{Xingang Chen,}
\author[c]{Mohammad Hossein Namjoo,}
\author[d, e]{Misao Sasaki,}
\author[f, g]{Dong-Gang Wang}
\author[a, h]{and Ziwei Wang}

\affiliation[a]{CAS Key  Laboratory for Researches in Galaxies and Cosmology, Department of Astronomy, School of Astronomy and Space Science, University of Science and Technology of China, Hefei, Anhui 230026, China}
\affiliation[b]{Institute for Theory and Computation,
Harvard-Smithsonian Center for Astrophysics, \\
Cambridge, MA 02138, USA}
\affiliation[c]{Department of Physics, Massachusetts Institute of Technology, Cambridge, Massachusetts 02139 USA}
\affiliation[d]{Center for Gravitational Physics, Yukawa Institute for
Theoretical Physics, Kyoto University,
Kyoto 606-8502, Japan}
\affiliation[e]{International Research Unit of Advanced Future Studies, Kyoto University, Japan}
\affiliation[f]{Leiden Observatory, Leiden University,
2300 RA Leiden, The Netherlands}
\affiliation[g]{Lorentz Institute for Theoretical Physics, Leiden University,\\
2333 CA Leiden, The Netherlands}
\affiliation[h]{Department of Physics, McGill University, Montreal, Quebec H3A 2T8, Canada}

\emailAdd{yifucai@ustc.edu.cn}
\emailAdd{xingang.chen@cfa.harvard.edu}
\emailAdd{namjoo@mit.edu}
\emailAdd{misao@yukawa.kyoto-u.ac.jp}
\emailAdd{wdgang@strw.leidenuniv.nl}
\emailAdd{ziwei.wang@mail.mcgill.ca}

\abstract{
Non-attractor inflation is known as the only single field inflationary scenario that can violate non-Gaussianity consistency relation with the Bunch-Davies vacuum state and generate large local non-Gaussianity.
However, it is also known that the non-attractor inflation by itself is incomplete and should be followed by a phase of slow-roll attractor.
Moreover, there is a transition process between these two phases. In the past literature, this transition was approximated as instant and the evolution of non-Gaussianity in this phase was not fully studied.
In this paper, we follow the detailed evolution of the non-Gaussianity through the transition phase into the slow-roll attractor phase, considering different types of transition.
We find that the transition process has important effect on the size of the local non-Gaussianity.
We first compute the net contribution of the non-Gaussianities at the end of inflation in canonical non-attractor models.
If the curvature perturbations keep evolving during the transition - such as in the case of smooth transition or some sharp transition scenarios - the $\mathcal{O}(1)$ local non-Gaussianity generated in the non-attractor phase can be completely erased by the subsequent evolution, although the consistency relation remains violated.
In extremal cases of sharp transition where the super-horizon modes freeze immediately right after the end of the non-attractor phase,  the original non-attractor result can be recovered.
We also study models with non-canonical kinetic terms, and find that the transition can typically contribute a suppression factor in the squeezed bispectrum, but the final local non-Gaussianity can still be made parametrically large.
}


\begin{document}

\maketitle
\flushbottom

\section{Introduction}

Inflationary cosmology is the leading paradigm of the very early universe \cite{Guth:1980zm, Linde:1981mu, Starobinsky:1980te, Brout:1977ix, Sato:1980yn, Fang:1980wi, Albrecht:1982wi}, in which the universe has experienced a primordial phase of quasi-de Sitter expansion. The simplest inflation model is realized by a canonical scalar field slowly rolling along a sufficiently flat potential. The associated perturbation theory successfully predicted a nearly scale-invariant power spectrum of primordial curvature perturbation, which is favoured by the latest cosmic microwave background (CMB) observations \cite{Ade:2015xua, Ade:2015lrj}. Moreover, it is widely acknowledged that the primordial non-Gaussianity, which encodes information about the very early universe, could be a powerful tool to discriminate different inflation models or alternative scenarios \cite{Bartolo:2004if,Liguori:2010hx,Chen:2010xka,Wang:2013eqj}.
Remarkably, there is a consistency relation for non-Gaussianity in single-field slow-roll inflation models pointed out by Maldacena \cite{Maldacena:2002vr, Creminelli:2004yq}. The consistency relation states that the amplitude of the primordial non-Gaussianity in squeezed configuration - where the wavelength of one mode is much larger than the other two in the three point correlation fucntion - is proportional to the spectral index of the power spectrum of scalar perturbations, {\it i.e.} $\fnl=5(1-n_s)/12$. Accordingly, the observation of the almost scale invariant power spectrum of linear perturbation indicates extremely small amount of nonlinear correlations in squeezed limit.
As a result, one expects that the simplest inflation model in terms of single slow-roll scalar field would be ruled out if any  squeezed limit non-Gaussianity  could be detected.

It is, however, interesting to notice that there exists a nontrivial inflationary scenario, dubbed as non-attractor inflation \cite{Kinney:2005vj, Namjoo:2012aa, Martin:2012pe, Chen:2013aj, Chen:2013eea, Huang:2013lda}, that can violate Maldecena's consistency relation even in the framework of single scalar field with Bunch-Davies initial states. This is due to the fact that curvature perturbations generated from quantum fluctuations during the non-attractor phase are dominated by the growing modes at super-Hubble scales, of which the behaviour is much similar to the matter bounce cosmology \cite{Wands:1998yp, Finelli:2001sr, Cai:2014bea} rather than the cosmology of slow-roll inflation. Accordingly, similar to the matter bounce cosmology \cite{Cai:2009fn, Li:2016xjb}, large amount of {\it local non-Gaussianity} - which contributes dominantly to the squeezed limit bispectrum -  can be achieved in non-attractor inflation models. Ref. \cite{Namjoo:2012aa} considers a simple model with canonical kinetic term which predicts $\fnl \simeq 5/2$. The idea is then further generalized to the models with non-canonical kinetic terms in \cite{Chen:2013aj, Chen:2013eea, Huang:2013lda} where it has been shown that the non-Gaussianity can be arbitrarily large. Inspired by this unconventional behaviour of primordial perturbations, many studies have been devoted to understand the possible violation of the consistency relation during the non-attractor phase from a variety of theoretical perspectives \cite{Mooij:2015yka, Pajer:2017hmb, Bravo:2017wyw, Finelli:2017fml}.

Furthermore, it is important to notice that the non-attractor inflation alone is not phenomenologically viable \cite{Namjoo:2012aa, Martin:2012pe, Cai:2016ngx}. Namely, without a conventional attractor phase, the non-attractor inflation does not provide enough e-folds or cannot fit the COBE normalization of the density perturbations.
For a more realistic consideration, the phase of non-attractor inflation shall be regarded as some initial stage of the whole inflationary era, and a phase transition from non-attractor to the slow-roll attractor evolution becomes essential for this class of models.
Therefore, the non-attractor inflation model consists of at least three different kinds of phases: the non-attractor phase, the transition phase, and the slow-roll phase. We shall define these phases more explicitly in models we study.
During the transition phase, modes that exited the horizon may not freeze, the main focus of this paper is to understand how the transition process would influence primordial non-Gaussianities generated in the non-attractor phase.

In this work, we revisit primordial non-Gaussianities from non-attractor inflation by focusing on the impact of the {\it non-attractor to attractor transition}.
We begin with a detailed analysis of the non-attractor inflation model with a canonical scalar field, which was previously studied in Ref.~\cite{Cai:2016ngx, Namjoo:2012aa}.
Here the transition processes are classified into two different cases, depending on whether the background evolution around the transition is smooth or sharp.
We first apply the in-in formalism to study the bispectrum in these two cases separately.
For the smooth transition, our calculation shows that the non-Gaussianity generated in the non-attractor phase {\it cannot} survive through the transition to the slow-roll attractor phase. So the value $\fnl=5/2$ generated during the non-attractor phase returns to $\sim 0$ (slow-roll-suppressed) in the slow-roll phase,
and the net contribution to the local $\fnl$ is negligible as in the slow-roll attractor case.
The situation is more complicated in the sharp transition. After a detailed analysis on the background and perturbations, we find that, in general the non-Gaussianity generated in the non-attractor phase is also suppressed after the transition. But in extremal cases where the curvature perturbations freeze out immediately at the transition time, the original result $\fnl\simeq5/2$ can be recovered.
We confirm all these results by employing the simple and intuitive calculation of the $\delta N$ formalism. Note that despite the non-trivial evolution of non-Gaussianity during the transition phase, the consistency relation is still violated even though the amplitude of non-Gaussianity might be slow-roll suppressed. This is a consequence of the fact that the curvature perturbation modes keep evolving after they crossed the Hubble horizon; in contrast with the conventional, slow-roll models where curvature perturbation is conserved on super-horizon scales.

We further study the transition process in the non-attractor inflation model driven by a non-canonical scalar field, as constructed in \cite{Chen:2013aj, Chen:2013eea}.
The background evolution shows that, the inflaton field first becomes canonical before the cosmological system enters into the phase of slow-roll attractor through a smooth transition phase.
The difference between these models and the above canonical model is that now we have two types of terms in the non-canonical models. The first type behaves very similarly to the interaction term in the canonical model, and it does not contribute to large local non-Gaussianity either when a smooth transition is taken into account. However, the non-canonical models have another set of qualitatively different terms. These second type of terms are unique due to the presence of the non-canonical kinetic terms. The contribution to large local non-Gaussianity from these terms do not get exactly erased by the smooth transition period, but instead gets an additional suppression factor.
Since the suppression factor and the amplitude of primordial non-Gaussianity generated in the non-attractor phase are independent of each other, the large local non-Gaussianity is still possible for certain model parameters. So the main conclusions of \cite{Chen:2013aj, Chen:2013eea} remain unchanged.

The paper is organized as follows.
In Section \ref{sec:canonical} we study the canonical model of non-attractor inflation.
After reviewing previous works, we focus on the detailed transition process from the initial non-attractor phase to the subsequent phase of slow-roll attractor.
Then we elaborate on the behaviour of local non-Gaussianity in two different cases -- smooth transition and sharp transition, via both in-in formalism and $\delta N$ formalism.
In Section \ref{sec:non-c} we generalize the study of the non-attractor inflation to models with non-canonical kinetic terms, where we only consider smooth transition case.
The detailed transition process in these models is shown by full analysis of the background dynamics.
After that, we estimate the size of the non-Gaussianity and find a suppression effect caused by the background evolution of the transition process.
We summarize our conclusions with a discussion in Section \ref{sec:concl}.
Throughout the paper we take the convention of the reduced Planck mass to be $\mpl^{2}=1/8\pi G=1$.

\section{The canonical model}
\label{sec:canonical}

In this section we revisit the calculation of primordial non-Gaussianities in the model of canonical non-attractor inflation,
and show how the different transition processes may change the non-Gaussianity generated in the non-attractor phase.

\subsection{The non-attractor phase and local non-Gaussianity}

The canonical non-attractor model is constructed by assuming that the inflaton's potential is almost a constant, i.e. for sufficiently large regime one has
$V(\phi) \simeq V_0$ \cite{Kinney:2005vj, Namjoo:2012aa}.
Accordingly, the background equations in this model are given by
\be \label{bgEoM}
\ddot\phi+3H\dot\phi \simeq 0~,~~~~ 3H^2 = \frac{1}{2}\dot\phi^2 + V \simeq V_0~,
\ee
where a dot denotes the derivative with respect to cosmic time $t$, and $H \equiv \dot a/a$ is the Hubble parameter.
This leads to the following behaviour for the slow-roll parameters
\be \label{srpara}
\epsilon \equiv -\frac{\dot H}{H^2} = \frac{\dot\phi^2}{2H^2} \propto a^{-6}~,~~~~ \eta \equiv \frac{\dot\epsilon}{H\epsilon} =-6~.
\ee
As shown in the above equation, the slow-roll parameter $\epsilon$ decays very quickly during the non-attractor phase, and thus, one can take the limit $\epsilon \rightarrow 0$ as a good approximation here.
As a result, the Hubble parameter $H$ is nearly constant during the non-attractor phase and in terms of conformal time $\tau$ the scale factor takes $a\simeq -1/(H\tau)$.
In addition, the second slow-roll parameter $\eta$ is of order $\mathcal{O}(1)$.

For the primordial curvature perturbation $\calR$, we define $z \equiv a\sqrt{2\epsilon}$ and $u_k \equiv z\calR_k$. Then at the linear level, the perturbation variable $u_k$ is governed by the Mukhanov-Sasaki equation
\begin{equation} \label{ms}
u_k'' + \( k^2 -\frac{z''}{z} \) u_k = 0~,
\end{equation}
where the prime denotes the derivative to conformal time $\tau$.
Following the standard treatment, the effective mass can be written as
$z''/z \simeq (\nu^2-1/4)/{\tau^2}$,
where for $\epsilon\ll1$, $\nu$ is given by
\begin{equation}
\label{nu}
\nu^2 = \frac{9}{4} +\frac{3}{2}\eta +\frac{1}{4}\eta^2+\frac{\dot\eta}{2H} +\mathcal{O}(\epsilon)~.
\end{equation}
In the non-attractor stage, $\eta=-6$, and thus, $\nu=3/2$.
Consequently, Eq. \eqref{ms} yields the mode function of curvature perturbation as follows,
\be \label{BD}
\calR_k(\tau)= \frac{u_k}{z} =\frac{H}{\sqrt{4\epsilon k^3}} (1+ik\tau){e^{-ik \tau}}~,
\ee
which looks the same as the one in the lowest order slow-roll approximation. But notice that $\epsilon$ is rapidly evolving here in contrary to the slow-roll case.
After Hubble-exit, one can get a scale-invariant power spectrum of primordial curvature perturbation, of which the form takes
$P_\calR(k)\equiv \frac{H^2}{8\pi^2\epsilon}$.
However, since $\epsilon\propto a^{-6}$, the amplitude of curvature perturbation grows as $\calR_k\propto a^{3}$ at super-Hubble scales.
As a result, the final form of the power spectrum ought to be evaluated after the end of the non-attractor phase.

In order to calculate the non-Gaussianity, one needs to study the three-point correlation function of primordial curvature perturbation
\be \label{bispectr}
\langle \calR_{\bf k_1} \calR_{\bf k_2} \calR_{\bf k_3} \rangle
\equiv (2\pi)^3 \delta^{(3)}({\bf k_1+k_2+k_3}) B_\calR(k_1, k_2, k_3)~.
\ee
At the squeezed limit $k_1\simeq k_2\gg k_3$, the bispectrum $B_\calR$ can be expressed as
\be
 B_\calR(k_1, k_2, k_3)=(2\pi)^4\frac{1}{k_1^3k_3^3} P_\calR(k_1) P_\calR(k_3)\frac{3}{5}\fnl ~,
\ee
where  $\fnl$ is the amplitude of non-Gaussianity in squeezed limit. The consistency relation, predicts $\fnl \simeq \frac{5}{12} (1-n_s)$ which we will see is violated in non-attractor models. Notice that the local shape has the same scaling behaviour in squeezed limit, although it is well defined in any configuration \cite{Chen:2010xka}. The non-Gaussianity that is generated during the non-attractor phase is indeed in the local shape but we are only interested in the squeezed limit (which tells us whether the consistency relation is violated or not); therefore we will not discuss non-Gaussianities in general configurations.

Ref.~\cite{Namjoo:2012aa} uses two methods to compute the size of local non-Gaussianity.
The first method focuses on the non-attractor phase alone.
Because the contributions from the terms in cubic Lagrangian are slow-roll suppressed in this phase, Ref.~\cite{Namjoo:2012aa} focuses on the contribution from a field-redefinition term in
\begin{equation}
\label{FR}
\calR = \calR_n +\frac{\eta}{4}\calR_n^2 + \frac{1}{H}\calR_n \dot\calR_n~,
\end{equation}
which yields
\be \label{fnl52}
\fnl=-\frac{5}{4}(\eta+4)=\frac{5}{2}
\ee
at the end of the non-attractor phase $\tau_e$.
If these perturbations got frozen immediately at the end of this phase and were carried along to the attractor slow-roll phase, we would end up with this order-one non-Gaussianity. However, the transition from the non-attractor phase to the slow-roll phase may not be an instant process and the process is not generically an attractor solution either. It turns out that the evolution of modes at the super-horizon scales can be non-negligible during this transition period.

The second method used in Ref.~\cite{Namjoo:2012aa} indeed considers this transition, but treating it as an instant process.
In this method, the field redefinition term no longer contributes because the parameter $\eta$ should now be evaluated at the end of inflation instead of at the end of the non-attractor phase. This value of $\eta$ is negligible.
The corresponding contribution should now, equivalently, come from an interaction term in the cubic Lagrangian,
\begin{equation}
\label{interact}
S_3 \supset \int dtd^3x \frac{a^3\epsilon}{2}\dot\eta \calR^2\dot{\calR}~.
\end{equation}
as correctly considered in Ref.~\cite{Namjoo:2012aa}.
The bispectrum coming from this interaction term is
\be \label{bispectinter}
B_\calR(k_1, k_2, k_3)
=-2{\Im}\calR_{{k}_1}(\tau_0)\calR_{{k}_2}(\tau_0)\calR_{{k}_3}(\tau_0) \int_{-\infty}^{\tau_0} d\tau{a^2\epsilon\eta'}\left[
\calR_{k_1}^*(\tau)
\calR_{k_2}^*(\tau)
\calR_{k_3}^{*\prime}(\tau)
+{\rm perm.}\right]~,
\ee
where  $\tau_0$ is {the conformal time after which the super-horizon curvature perturbation as well as the corresponding bispectrum cease evolving}. We also remind that $\tau_e$ denotes the end of the non-attractor phase.
The $\eta$ parameter goes from $-6$ to nearly zero and then the coefficient $\eta'$ can be comparably large.
If the transition is approximated as {an instant process} that takes place suddenly at the time $\tau_e$ when the non-attractor phase ends \cite{Namjoo:2012aa}, then {one may expect} $\tau_0=\tau_e$ and the behaviour of $\eta$ during the transition period can be approximated by a step function
\be \label{parametr}
\eta=-6\[1-\theta(\tau-\tau_e)\]~.
\ee
As a result, the interaction term \eqref{interact} leads to
\be \label{naive}
\lim_{k_3/k_1 \rightarrow 0} B_\calR(k_1, k_2, k_3)=(2\pi)^4\frac{1}{4k_1^3k_3^3} P_\calR(k_1) P_\calR(k_3)
\int d\tau \eta' ~,
\ee
and the value $\fnl=5/2$ will be recovered.
However, one may still wonder whether this conclusion holds true if we consider a complete transition process.
In the next subsections, we will study various transition cases in details, and show that, for a smooth transition the actual contribution from \eqref{interact} is negligible; while the $\mathcal{O}(1)$ local non-Gaussianity can be recovered from a sharp transition.

\subsection{The non-attractor to slow-roll transitions}
\label{sec:transition}

The reason that the ultra-slow-roll inflation with a constant potential cannot be a complete model (even if we impose an abrupt cutoff and start the reheating instantly in the non-attractor phase) is that, after $40\sim 60$ efolds, the density perturbation cannot produce the observed value.\footnote{If we require only the non-attractor inflation to solve the flatness and horizon problems, the total number of efolds of the non-attractor phase should be $40\sim 60$ efolds, at the end of which the value of $\epsilon$ would be diminishingly small. To fit the COBE normalization, $H$ would be diminishingly small and ruled out already.} So a transition to a slow-roll phase is needed.
In the following, we construct a model that describes such a transition. The advantage of our model is that the exact analytical solutions can be obtained, in which the inflaton field begins the evolution in the non-attractor phase and then joins the slow-roll phase gradually.

\begin{figure}[tbhp]
\centering
\includegraphics[width=0.5\linewidth]{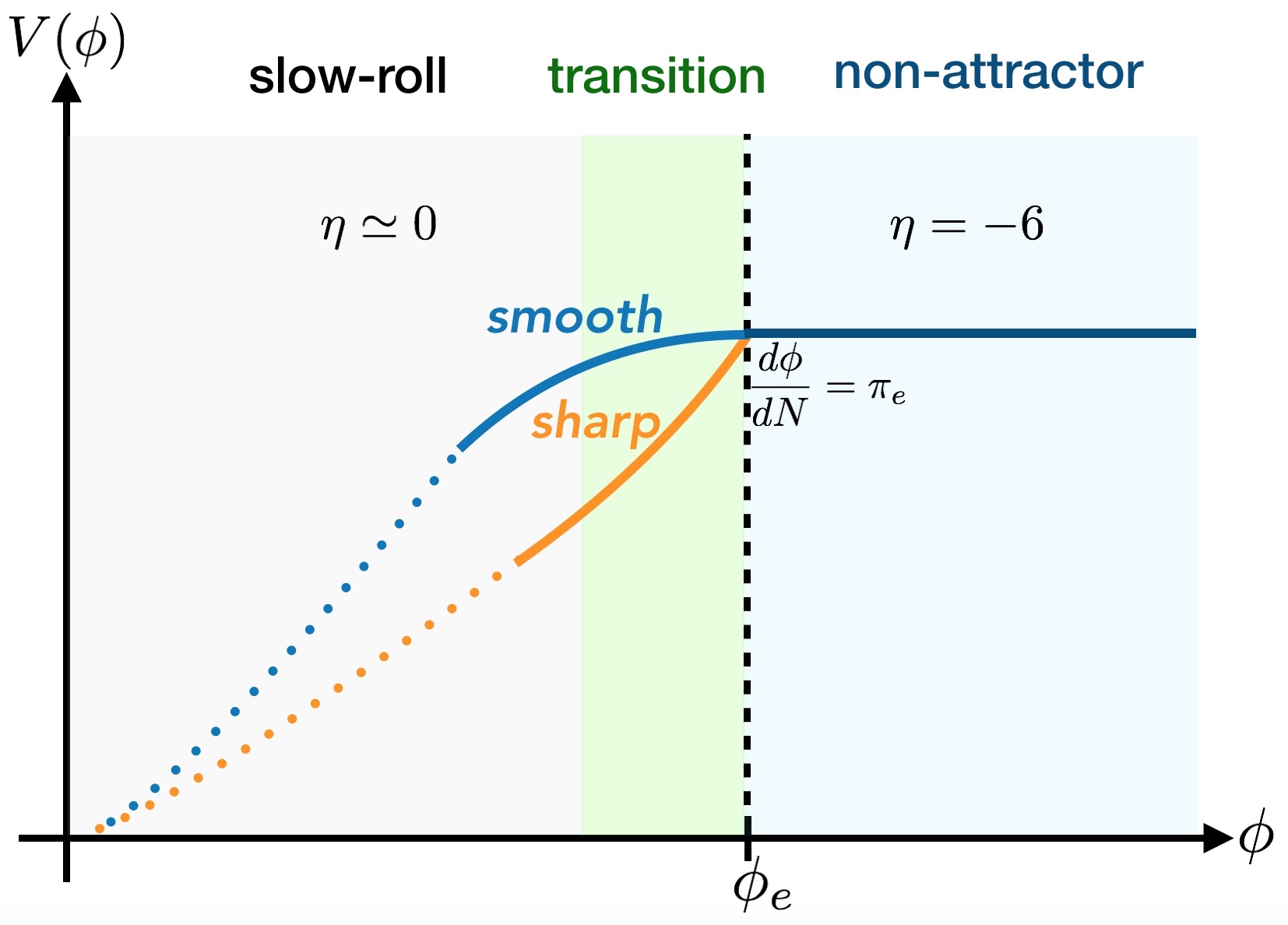}
\caption{A sketch plot of the potentials of non-attractor inflation with smooth and sharp transitions. Note that the inflaton rolls from right to left, i.e. $\phi$ is decreasing during the evolution.}
\label{fig:potential}
\end{figure}

Suppose that the non-attractor phase ends at $\phi_e=\phi(\tau_e)$, and after that, a slow-roll potential $V(\phi)$  is attached to the constant one.
Since the transition process is very short and the inflaton field excursion is very tiny, during this period, the attached slow-roll potential can be expanded as follows,
\be \label{potential}
 V(\phi)=V(\phi_e)+\sqrt{2\epsilon_V}V(\phi_e)(\phi-\phi_e)+\frac12\eta_VV(\phi_e)(\phi-\phi_e)^2+\dots ~.
\ee
Here we have introduced the potential slow-roll parameters $\epsilon_V\equiv {1}/{2}({V'(\phi_e)}/{V(\phi_e)})^2$  and $\eta_V\equiv {V''(\phi_e)}/{V(\phi_e)}$,
which are expected to be small constants such that the slow-roll dynamics can be triggered after the transition.
Accordingly, one can sketch the possible form of potentials depending on different values of parameters which correspond to different types of transition, as shown in Figure \ref{fig:potential}.
We may distinguish two extreme possibilities:
if we require the derivative of the potential to be continuous, then $\epsilon_V = 0$ and thus the transition is {\it smooth};
{whereas for other cases, such as $\sqrt{2\epsilon_V}\gtrsim|\eta_V|$}, we get {\it sharp} transition. Note that by considering the above potential we restricted ourselves to the case with continuous potential and the positivity of the second term also implies that the inflaton rolls-down instead of jumping up.
By the end of this section, however, we will discuss how the results may change by considering non-standard cases of discontinuous potential or negative slope.
Finally, notice that the above additional potential in a single field model of inflation, breaks the internal shift symmetry explicitly; therefore even the generalized consistency relations \cite{Finelli:2017fml,Bravo:2017wyw} are not applicable, unless if the bispectrum does not evolve when the potential \eqref{potential} switches on.

In this type of inflation model, initially the inflaton field rolls along the constant potential $V=V_0$ for $\phi>\phi_e$, which we define as the {\em non-attractor phase}.
After the inflaton field reaches $\phi_e$, the potential becomes \eqref{potential}, on which inflation transits to the slow-roll attractor. We define this period as the {\em transition phase}, as shown by the light green region in Figure \ref{fig:potential}.
Using e-folding number $N$ as variable (with the convention $dN=H dt$), the background equations  become
\be
\frac{d^2\phi}{d N^2}+3\frac{d\phi}{d N} + 3\sqrt{2\epsilon_V} +3\eta_V(\phi-\phi_e) \simeq 0~,~~~~{\rm and}~~~~3H^2\simeq V(\phi_e) ~,
\ee
where we have assumed that the Hubble parameter is a constant. Without losing generality, we can set $N=0$ at $\phi_e$ and the field velocity at the same moment is introduced to be $\pi_e$, then we have the following analytical solution
\be \label{anasol1}
\phi =\frac{s-3-h}{s(s-3)}\pi_e e^{\frac{1}{2}(s-3)N} -\frac{s+3+h}{s(s+3)}\pi_e e^{-\frac{1}{2}(s+3)N}
+\frac{2\pi_eh}{s^2-9} +\phi_e ~,
\ee
\be \label{anasol2}
\pi\equiv\frac{d\phi}{d N}=e^{-3N/2}\[\pi_e\cosh\(\frac{s}{2}N\)
-\frac{{3+h}}{s} \sinh\(\frac{s}{2}N\) \] ~,
\ee
with the parameters
\be
 s\equiv \sqrt{9-12\eta_V}\simeq 3-2\eta_V ~,~~ h\equiv6 \sqrt{2\epsilon_V}/\pi_e ~,
\ee
being introduced. Notice that in our convention $\pi_e <0$ (because $\phi$ is decreasing throughout the evolution) and hence $h<0$. After some simple algebra, the slow-roll parameters defined in \eqref{srpara} during the transition are given by
\be \label{epstr}
\epsilon(N) = \frac{\pi_e^2}{2} e^{-3N}\[ \cosh\(\frac{s}{2}N\)
-\frac{{3+h}}{s} \sinh\(\frac{s}{2}N\) \]^2~,
\ee
\be \label{etatr}
\eta(N)=s-3-\frac{2s( 3+s +h )}{e^{sN} (s-3-h)+3+s+h}~.
\ee
We can see from the above that, as $N$ increases, $\eta$ goes from $-6-h$ to $-2\eta_V$ during the transition.

\begin{figure}[tbhp]
\centering
\includegraphics[width=0.475\linewidth]{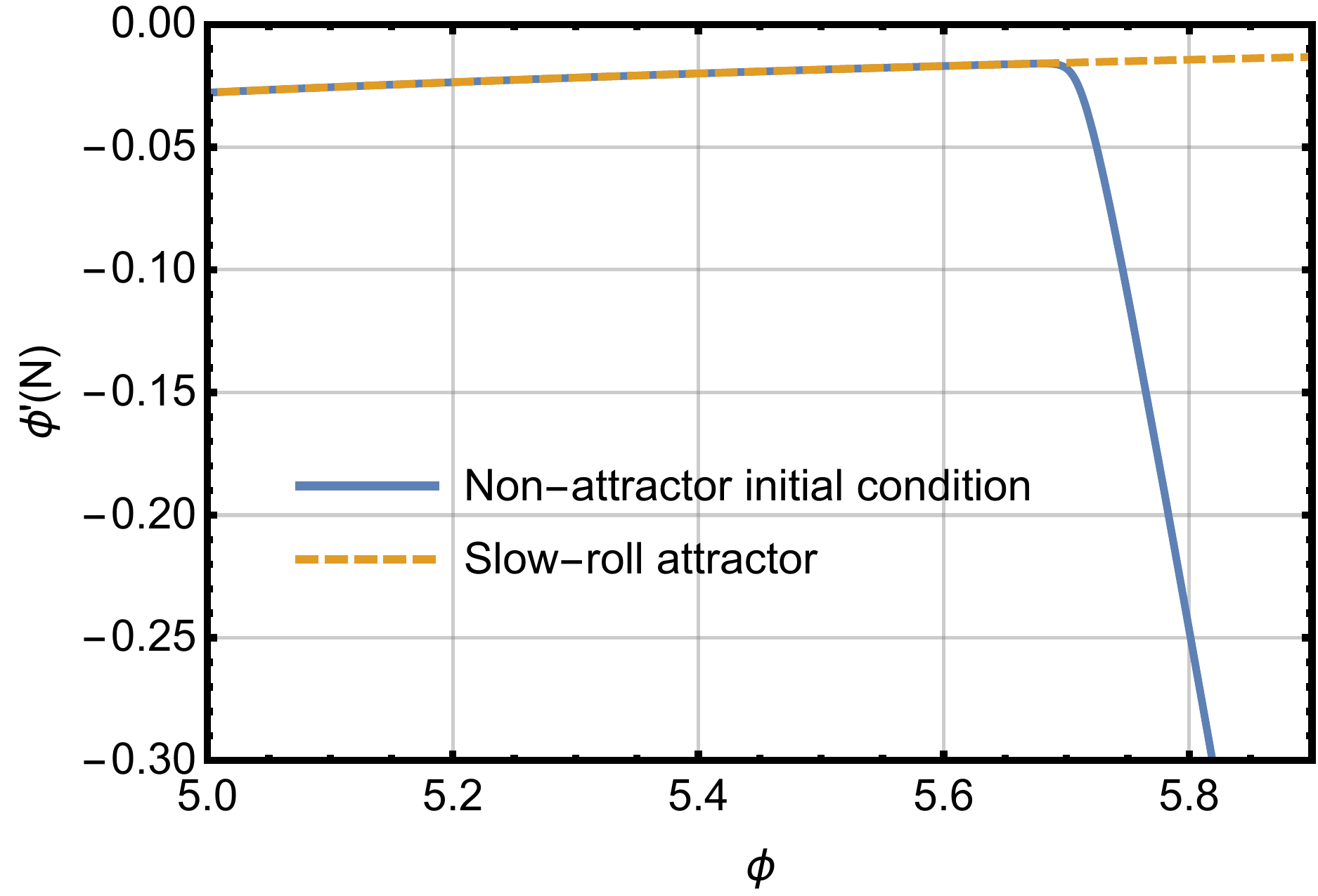}
\includegraphics[width=0.45\linewidth]{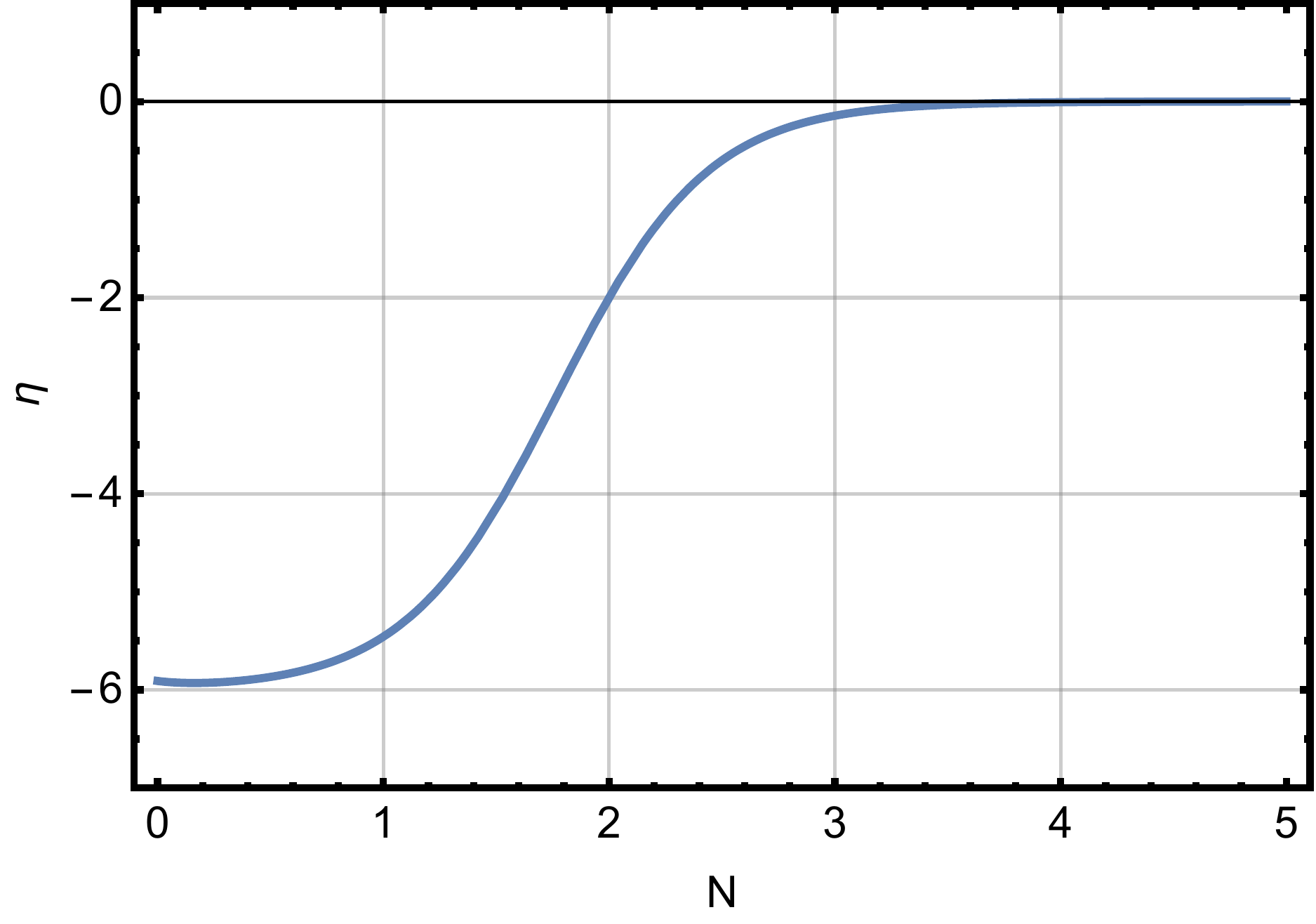}
\caption{Smooth transition. Left Panel: the phase space diagram of non-attractor to slow-roll transition on a plateau-like potential. Right Panel: the evolution of $\eta$ parameter during the transition.}
\label{srpara2}
\end{figure}

Note that, the background evolutions in smooth and sharp transitions behave manifestly different,
and $h$ is a crucial parameter to characterize their difference.
For the smooth transition, $h\to 0$, and thus at the beginning of the transition phase $\eta = -6$, which continuously follows the non-attractor phase and then smoothly evolves to the slow-roll attractor.
Figure \ref{srpara2} shows this behaviour via the phase space diagram and the evolution of $\eta$, where the smooth transition is depicted by the numerical solution of the non-attractor initial condition on a plateau-like potential\footnote{See Section \ref{sec:numeric} for more discussions about this implementation.}.

\begin{figure}[tbhp]
\centering
\includegraphics[width=0.475\linewidth]{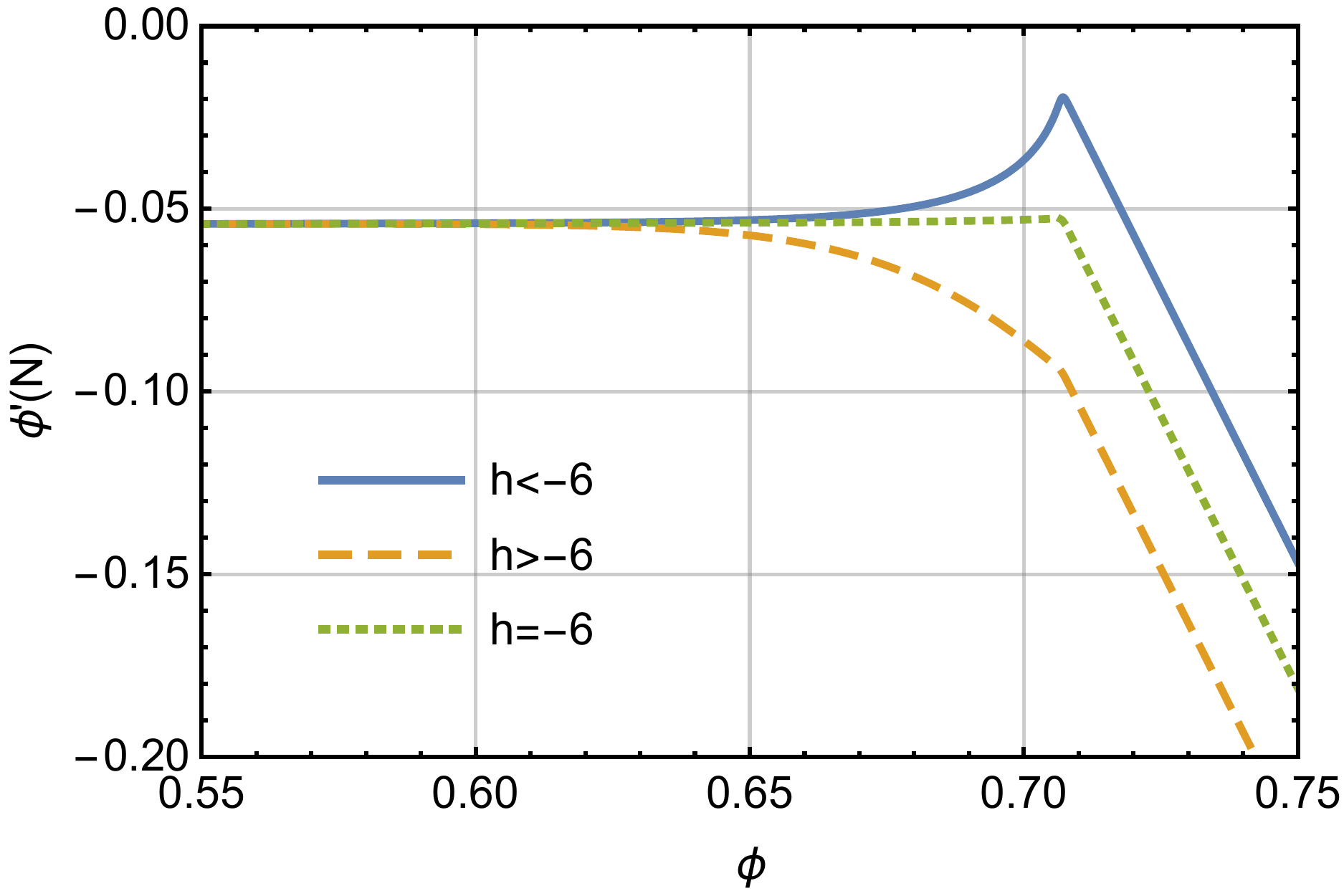}
\includegraphics[width=0.475\linewidth]{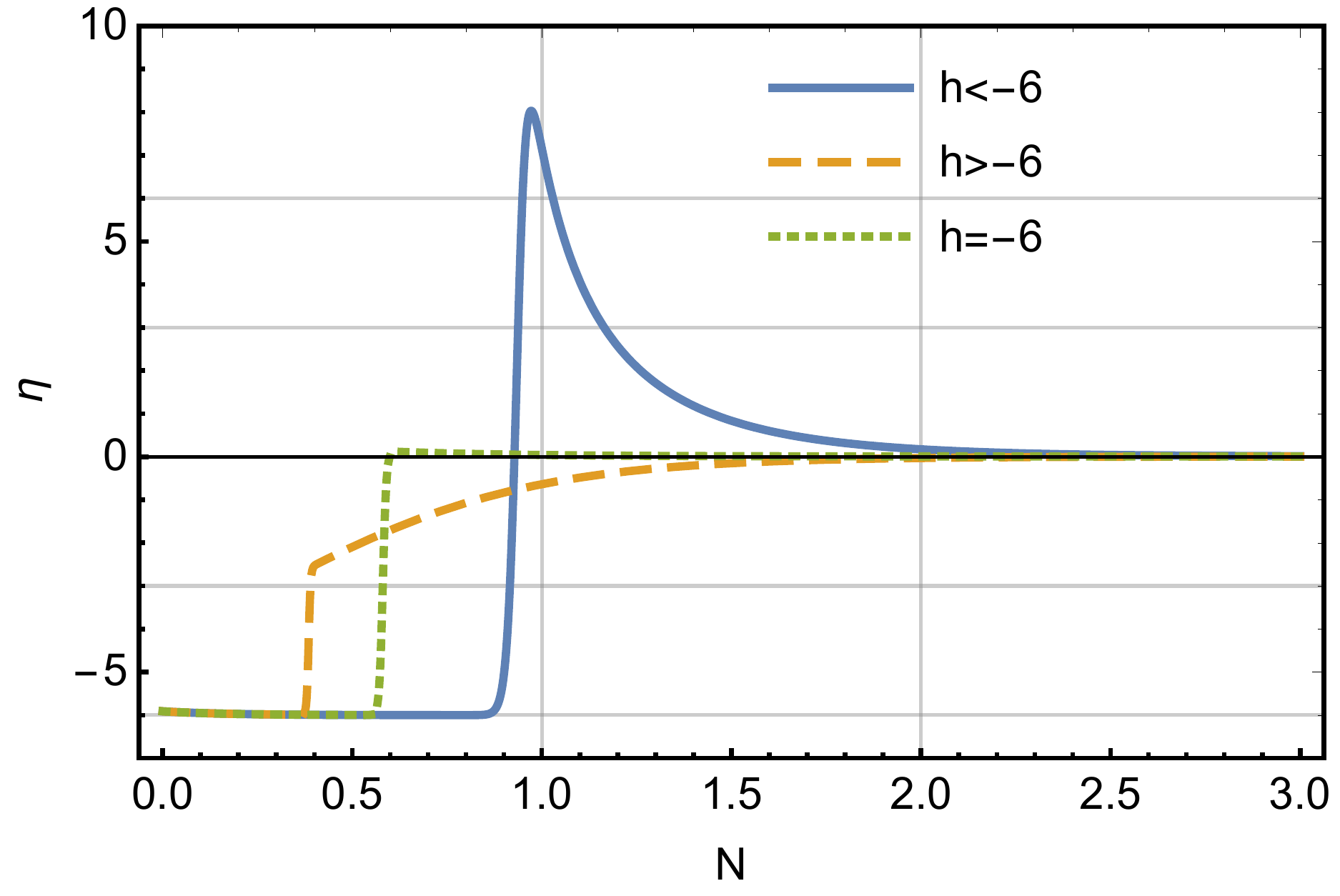}
\caption{Sharp transition. Left Panel: the phase space diagram of sharp transition for three different cases. Right Panel: the evolution of $\eta$ parameter during the sharp transition.}
\label{fig:sharp}
\end{figure}

For the sharp transition, $h$ is a negative constant determined by the field velocity $\pi_e$ at the end of the non-attractor phase.
From \eqref{epstr} we see that, when the attractor is reached after the sharp transition, we have $\epsilon_0\simeq \epsilon_V$
with $\epsilon_0=\pi_0^2/2$, where $\pi_0$ is the field velocity $\frac{d\phi}{dN}$ during the slow-roll phase. Therefore, the parameter $h$ can be described also by the ratio between $\pi_0$ and $\pi_e$
\be
h\equiv6\sqrt{2\epsilon_V}/\pi_e \simeq 6\sqrt{2\epsilon_0}/\pi_e=-6\pi_0/\pi_e ~.
\ee
From the relative magnitudes of $\pi_e$ and $\pi_0$, {it is straightforward to see that, the value of $|h|$ can be of order unity or even bigger, and
 there are three possible cases in sharp transition: $h<-6$, $h=-6$ and $-6<h<0$, as shown in the phase space diagram in Figure~\ref{fig:sharp}. Consequently
at the beginning of the transition $\eta=-6-h$ can be quite large, which differs from its value during the non-attractor phase (where it is $\eta=-6$).
Thus there is a sudden change of $\eta$ at the transition time, from $-6$ to $-6-h$, as shown by the numerical examples in the right panel of Figure~\ref{fig:sharp}.
For the later convenience, we formulate the evolution of $\eta$ around $\tau_e$ as
\be \label{step}
\eta=-6-h \theta(\tau-\tau_e)~,~~~~\tau_{e_-}<\tau<\tau_{e_+}~.
\ee
Therefore, typically a sharp transition process consists of an instant transition at the beginning and a following period of relaxation described by \eqref{anasol1} -- \eqref{etatr}.
One special case is $h=-6+2\eta_V \simeq -6$, where inflaton joins the slow-roll attractor immediately after the instant transition
 and there is no relaxation process.
However, it still differs from the oversimplified case in \eqref{parametr}. As we shall show in Section \ref{sec:sharp}, this realistic instant transition does not imply immediate freezing of the curvature perturbation (i.e. $\tau_0\neq\tau_e$), and the evolving super-horizon mode after the instant transition can still modify the non-Gaussianity generated during the non-attractor phase.}

With these background solutions of transitions, in the following we shall perform a detailed study of non-Gaussianities.
The in-in formalism is applied in Section \ref{sec:smooth} and \ref{sec:sharp},
for smooth and sharp transitions respectively.
In Section \ref{sec:deltaN}, we further confirm the in-in results in both cases via $\delta N$ formalism.

\subsection{Non-Gaussianity in a smooth transition}
\label{sec:smooth}

In this subsestion, we focus on in-in calculation of the smooth transition case, which corresponds to the limit $\epsilon_V \to 0$ in the potential \eqref{potential}, and demonstrate that there is a cancellation for the local non-Gaussianity generated during the non-attractor stage.
Then in Section \ref{sec:numeric}, we confirm this conclusion by the numerical study of a realistic model.
At last, in Section \ref{sec:general}, we perform an extended analysis to show that this conclusion holds true for smooth transition in general.

Before the in-in calculation, we should first check the behaviour of the mode function during the transition, which is governed by the Mukhanov-Sasaki equation \eqref{ms} and the index $\nu$ in \eqref{nu}.
Even though $\eta$ and $\dot\eta$ varies dramatically during the transition, surprisingly the exact solution \eqref{epstr} and \eqref{etatr} gives us $\nu^2=9/4-3\eta_V$, which is constant and the same as the result in slow-roll attractors\footnote{In Section \ref{sec:general}, we shall show that the cancellation giving this result of $\nu^2$ is not a coincidence.}. Therefore, the mode function in \eqref{BD} still applies here as the leading order approximation, and the resulting power spectrum in this period is still nearly scale-invariant.
We should further remark that, the curvature perturbation still evolves during the transition, and should be fixed after the slow-roll attractor is reached.
That is to say the final amplitude of the power spectrum is $P_\calR(k)\equiv \frac{H^2}{8\pi^2\epsilon_0}$, where $\epsilon_0$ is the $\epsilon$ in the slow-roll stage.

With this analytical description of the smooth transition, now let us look at the bispectrum caused by the cubic interaction term \eqref{interact}.
We can substitute the mode function \eqref{BD} into the in-in integral in \eqref{bispectinter}.
Notice that,
even though $\epsilon$ is small, it varies fast during the transition, thus
\be
\calR_k'(\tau) =\frac{H}{\sqrt{4\epsilon k^3}} k^2\tau{e^{-ik \tau}}-\frac{\eta}{2}aH\frac{H}{\sqrt{4\epsilon k^3}} (1+ik\tau){e^{-ik \tau}}~,
\ee
where the second term is due to the $\epsilon$'s evolution\footnote{This contribution was neglected in the calculation of \cite{Cai:2016ngx}, see Eq.(4.15) there.}.
Taking the squeezed limit $k_1=k_2=k\gg k_3$, we get
\be \label{bispecfull}
B_\calR(k_1, k_2, k_3)=-\frac{(2\pi)^4}{4k_1^3k_3^3} P_\calR^2  {\Im}
\int_{-\infty}^{\tau_0}  d\tau\frac{\eta'}{\sqrt{\epsilon/\epsilon_0}}e^{2ik(\tau-\tau_0)}
\[\frac{1-ik\tau}{k\tau}+\frac{3\eta}{4}\frac{(1-ik\tau)^2}{k^3\tau^3}\]
(1+ik\tau_0)^2~.
\ee
Since $\eta'$ is negligible during the non-attractor and slow-roll phase, we only need to compute this integral during the transition process (from $\tau_e$ to $\tau_0$).
As mentioned earlier, the evolution of the bispectrum after $\tau_0$ is suppressed, as is well-known in the attractor case where the super-horizon curvature perturbation freezes out.
Since we are mainly interested in the perturbation modes which exit the Hubble radius during the non-attractor phase, we can use $|k\tau_e|<|k\tau_0|\ll1 $. Thus the leading order contribution of the above bispectrum becomes
\be \label{bispectr2}
B_\calR(k_1, k_2, k_3)=-\frac{(2\pi)^4}{4k_1^3k_3^3} P_\calR^2 \int_{\tau_e}^{\tau_0} d\tau\frac{\eta'}{\sqrt{\epsilon/\epsilon_0}} \[1+\frac{\eta}{2}-\frac{\eta}{2}\(\frac{\tau_0}{\tau}\)^3\] ~.
\ee
Plugging in the analytical expressions for $\epsilon$ and $\eta$ in  \eqref{epstr} and \eqref{etatr}, we find after the transition
\be
\frac{3}{5}\fnl\simeq -\frac{\sqrt{2\epsilon_0}}{\pi_e} \frac{\eta_V}{2}~.
\ee
Here $\sqrt{2\epsilon_0}$ can be expressed as the field velocity $\frac{d\phi}{dN}$ at the beginning of the slow-roll phase $\tau_0$, thus ${\sqrt{2\epsilon_0}}\ll|{\pi_e}|$. As a result, the local non-Gaussianity becomes negligible after the transition.

If we compare this calculation with the result \eqref{naive} in the instant transition approximation, we can just identify $\tau=\tau_0$ and $\epsilon =\epsilon_0$ in \eqref{bispectr2} using the step function \eqref{parametr} for $\eta$.
However, here when we compute the smooth transition explicitly, the third term in the bracket becomes negligible, since $\tau_0/\tau<1$ during the transition. And we have seen that there is a cancellation between the first two terms, in contrast with the instant transition approximation which gives order one result. This cancellation is also demonstrated numerically as follows.

\subsubsection{Numerical study on a plateau-like potential}
\label{sec:numeric}

In a realistic case, non-attractor inflation can be seen as imposing the ultra-slow-roll initial condition on a plateau-like inflaton potential (such as Starobinsky inflation \cite{Starobinsky:1980te} and $\alpha$-attractors \cite{Kallosh:2013hoa, Kallosh:2013yoa}).
In such a situation, the smooth transition to slow-roll attractor occurs automatically.
In addition, due to the scale-invariant power spectrum generated in the initial non-attractor phase,
 the primordial perturbations can be suppressed on large scales, which is favored by current CMB observations \cite{Cai:2016ngx}.

Now we study the background evolution of this realistic model numerically, and then further check the analytical results above.
Consider the following potential of Starobinsky inflation \cite{Starobinsky:1980te}
\be
V(\phi)=V_0\(1-e^{-\sqrt{2/3}\phi}\)^2~,
\ee
which is very flat for large $\phi$. In the slow-roll attractor, the field velocity satisfies $\dot\phi_{sr}=-V'/(3H)$.
However, if inflation starts with a much larger velocity $|\dot\phi|\gg|\dot\phi_{sr}|$ on this very flat potential, initially it would be in the non-attractor phase.
Solving the background equations numerically, we get the results shown in
Figure \ref{srpara2}.
As we can see from the phase space diagram and the evolution of $\eta$, this realistic model indeed has a non-attractor initial phase,
and then it will join the slow-roll attractor very quickly.

\begin{figure}[tbhp]
\centering
\includegraphics[width=0.6\linewidth]{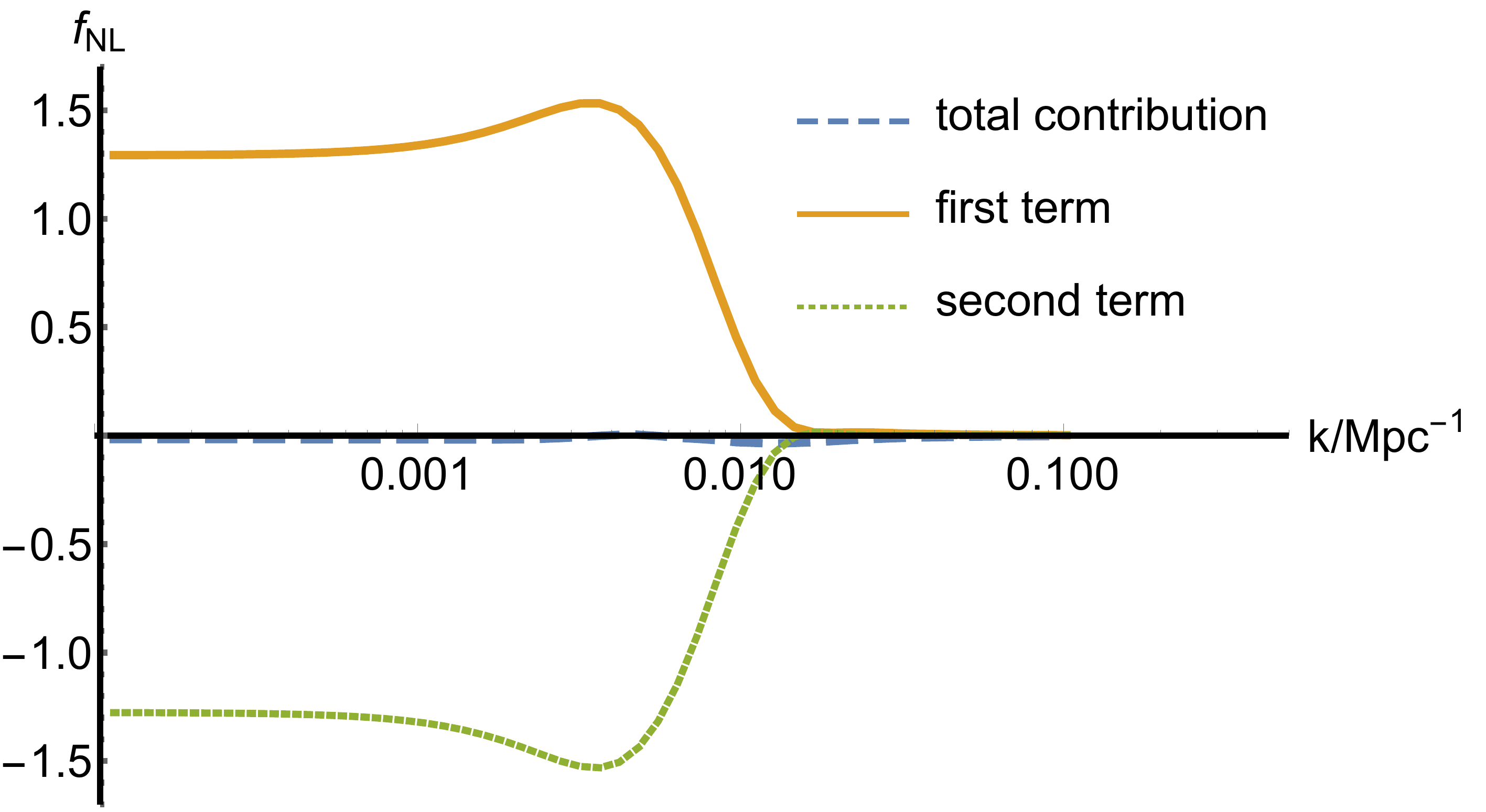}
\caption{The cancellation in the in-in integral \eqref{bispecfull}. }
\label{cancel}
\end{figure}

With this numerical solution, we can go back to do the full computation for the integral in \eqref{bispecfull}, not only for the perturbation which exit the Hubble radius before the transition (non-attractor modes), but also for those small scale modes (slow-roll modes).
The final bispectrum receives contributions from both terms in \eqref{bispecfull}.
The numerical result of local $\fnl$ as a function of $k$ is shown in Figure~\ref{cancel}.
As we see, if we only consider one contribution, the local non-Gaussianity is $\mathcal{O}(1)$ for the non-attractor modes, and then it vanishes for the slow-roll modes.
However, when we combine these two contributions together, they cancel each other and yield vanishing $\fnl$ even for non-attractor modes.
This result confirms the analytical calculation above.

In summary, both analytical and numerical calculation of the smooth transition process show that, there is a mysterious cancellation happening during this transition period. In the following subsection, we shall understand this cancellation in a more general way.

\subsubsection{A more general analysis}
\label{sec:general}

To understand what is going on during a smooth transition, we present a more general analysis as follows. First of all, let us remind of the background equations
\begin{equation} \label{bgeqs}
	\ddot\phi +3 H \dot\phi +V'(\phi) = 0~,~~~~3H^2 = \frac{1}{2}\dot\phi^2 + V(\phi) ~,
\end{equation}
Here the potential is required to have a slow-roll attractor,
but for now we do not assume any slow-roll conditions.
Then using the background equations and the Hubble "slow-roll" parameters defined in \eqref{srpara}, the second and third order derivatives of the slow-roll potential can be exactly expressed as
\ba
&&V''=\(6\epsilon -\frac{3}{2}\eta-\frac{\eta^2}{4}+\frac{5}{2}\epsilon\eta -2\epsilon^2-\frac{\dot\eta}{2H}\)H^2~,\nn\\
&&V'''=\frac{1}{\sqrt{2\epsilon}}\(9\epsilon\eta -\frac{3\dot\eta}{2H}-\frac{\eta\dot\eta}{2H}+3\epsilon\eta^2+\frac{3\epsilon\dot\eta}{H} -9\epsilon^2\eta-\frac{\ddot\eta}{2H^2}-12\epsilon^2+4\epsilon^3\)H^2~,
\ea
which respectively correspond to the inflaton mass and self-coupling.
Note that these derivatives of the potential should be suppressed so that the slow-roll attractor is possible. Due to this requirement, some useful combinations of $\eta$ and $\dot\eta$, that we will soon encounter, should be much smaller than unity, even though $\eta$  and $\dot\eta$ can be individually large during the non-attractor and transition stages.
One consequence of this observation is the behaviour of the effective mass in the Mukhanov-Sasaki equation \eqref{ms}.
As we mentioned in the last subsection, the coefficient $\nu^2-9/4$ there is always small, even during the transition where $\eta$ and $\dot\eta$ are big.
Now we see this parameter is directly related to inflaton mass
\begin{equation}
\nu^2-\frac94= -\frac{V''}{H^2} +\mathcal{O}(\epsilon)~,
\end{equation}
which does not care if inflation is in the attractor or not.

With this knowledge, let us look at the cubic interaction term \eqref{interact} again.
In our in-in calculation above, one subtlety is caused by the evolution behaviour of $\dot\calR$.
We can remove it via integration by part, and express \eqref{interact} as
\begin{equation}
 -\int dtd^3x \frac{d}{dt}\(\frac{a^3\epsilon\dot\eta}{6}\) \calR^3 + {\rm surface~ term} ~.
\end{equation}
Since there is no more time derivative on $\calR$, the only important effect lies in the cubic coupling.
Here we are encouraged to introduce the effective coupling as
\be
\frac{1}{6a^3\epsilon}\frac{d}{dt}\({a^3\epsilon\dot\eta}\)=\frac{H^2}{3}
\(\frac{3\dot\eta}{2H}  + \frac{\eta\dot\eta}{2H}  + \frac{\ddot\eta}{2H^2} \)~.
\ee
Again it looks like due to the drastic variation of $\eta$, these terms could be large during the transition. However, if we plug in our analytical and numerical solutions in the last section, this combination is shown to be negligible.
Interestingly, they are also present in $V'''$, and can be written as
\be
\frac{1}{6a^3\epsilon}\frac{d}{dt}\({a^3\epsilon\dot\eta}\)=-
\frac{1}{3} \sqrt{2\epsilon} V''' + \mathcal{O}(\epsilon)H^2~.
\ee
Therefore the contribution from this term is always small, no matter how big $\eta$ and $\dot\eta$ are during the transition.
The presence of $V'''$ is not a coincidence here. In the flat gauge, the operator which contributes to the cubic Lagrangian \eqref{interact} comes from the self-interaction of field fluctuations
\be
\mathcal{L}_3\subset\frac{a^3}{6}V'''\delta\phi^3 = \frac{a^3\epsilon}{3}\sqrt{2\epsilon}V'''\calR^3~.
\ee
Taking the decoupling limit, we have
\be
V'''=\frac{1}{\sqrt{2\epsilon}}\(-\frac{3\dot\eta}{2H}-\frac{\eta\dot\eta}{2H}-\frac{\ddot\eta}{2H^2}+ \mathcal{O}(\epsilon)\)H^2~.
\ee
And after integration by parts, the self-interaction term exactly gives us the cubic term \eqref{interact}.

In summary, for a smooth non-attractor to slow-roll transition,
as long as we have a slow-roll potential ($V'''$ is small), the non-Gaussianities would be always small.
The cubic interaction term \eqref{interact}, which was previously thought to contribute sizable $\fnl$ in the instant transition approximation, actually never contributes in the realistic smooth transition case.
However, this argument may not work in sharp transition cases. There the potential is unsmooth around the transition, which may yield large $V'''$. Furthermore, the unconventional behaviour of the mode function will add extra complications.
These issues of the sharp transition will be addressed in the next subsection.

\subsection{Non-Gaussianity in a sharp transition}
\label{sec:sharp}

As we discussed previously, the background of sharp transition differs from the smooth transition case.
Now we come to study the effect of a sharp transition on the evolution of perturbations, especially on the local non-Gaussianity. The sharp transition corresponds to the case where the second term in the potential \eqref{potential} {is also important.
In this subsection, we shall study the case $\sqrt{2\epsilon_V}\gtrsim|\eta_V|$.
The form of the potential \eqref{potential} can be invalid after the inflaton field evolves to sufficiently large distances from the transition point.}
But we can assume that the slow-roll limit is already reached before that happens so that we do not need to keep track of perturbations any more.

First of all, unlike the smooth transition case, the behaviour of the mode function in the sharp transition is more complicated.
If we look at the Mukhanov-Sasaki equation and the index $\nu$ in \eqref{nu},
the analytical solution of the sharp transition \eqref{epstr} and \eqref{etatr} still gives us $\nu^2=9/4-3\eta_V$.
However, due to the sudden change of $\eta$ at the transition time $\tau_e$, one cannot simply continue using the initial mode function \eqref{BD} after $\tau_e$.
Here when the transition happens, the matching condition requires the mode function and its first derivative to be continuous, i.e. $\calR(\tau_{e_-})=\calR(\tau_{e_+})$ and $\calR'(\tau_{e_-})=\calR'(\tau_{e_+})$.
This gives us the following behaviour of curvature perturbation after $\tau_{e}$
\be \label{nonBD}
\calR_k(\tau)= \alpha_k \frac{H}{\sqrt{4\epsilon k^3}} (1+ik\tau){e^{-ik \tau}}
+\beta_k \frac{H}{\sqrt{4\epsilon k^3}} (1-ik\tau){e^{ik \tau}}~,
\ee
\ba \label{nonBD2}
\calR_k'(\tau) &=& \alpha_k \[\frac{H}{\sqrt{4\epsilon k^3}} k^2\tau{e^{-ik \tau}} + \frac{\eta}{2\tau} \frac{H}{\sqrt{4\epsilon k^3}} (1+ik\tau){e^{-ik \tau}} \] \nn\\
&&+ \beta_k \[\frac{H}{\sqrt{4\epsilon k^3}} k^2\tau{e^{ik \tau}} + \frac{\eta}{2\tau} \frac{H}{\sqrt{4\epsilon k^3}} (1-ik\tau){e^{ik \tau}} \]~,
\ea
where
\be
\alpha_k = 1+i\frac{h}{4k^3\tau_e^3}(1+k^2\tau_e^2)~,~~~~
\beta_k=-ih(1+ik\tau_e)^2\frac{e^{-2ik\tau_e}}{4k^3\tau_e^3}~.
\ee
We can easily check the long wavelength behaviour of the mode function after $\tau_e$
\be
\label{kto0_mode}
\calR_k(\tau) \simeq  \frac{6-h}{6}\frac{H}{\sqrt{4\epsilon k^3}} + \frac{\tau^3}{6\tau_e^3}\frac{H}{\sqrt{4\epsilon k^3}}  ~~~~ \text{ for} ~ k\rightarrow 0 ~.
\ee
This solution satisfies the super-horizon EoM: $\ddot\calR+(3+\eta)H\dot\calR=0$. At the time $\tau_0$ of the slow-roll stage, we get the freezed amplitude
\be
\calR_k(\tau_0) \simeq  \frac{6-h}{6}\frac{H}{\sqrt{4\epsilon_0 k^3}} = \( 1+\sqrt{\frac{\epsilon_0}{\epsilon_e}} \)\frac{H}{\sqrt{4\epsilon_0 k^3}} ~ .
\ee
For  large values of  $|h|$ the above relation reduces to $\calR_k(\tau_0) \simeq \frac{H}{\sqrt{4\epsilon_e k^3}}$
which is similar to the mode function at the transition time $\tau_e$. Thus, for $|h| \gg 1$, the final power spectrum does not change much by the transition and we have $P_\calR\simeq\frac{H^2}{8\pi^2\epsilon_e}$. Therefore, we expect to recover the previously calculated non-Gaussianity $\fnl=5/2$ in the $|h| \gg 1$ limit where the mode function is assumed to freeze instantly after transition. We will confirm this expectation explicitly below.  Note also that, {in the $h=-6$ case with only instant transition, the super-horizon modes still evolve from $\tau_e$ to $\tau_0$, as can be seen from \eqref{kto0_mode}.
This shows that a realistic instant transition to the slow-roll evolution (which corresponds to $h=-6$) does not imply an instant freezing of the mode function, thus we do not expect to recover $\fnl=5/2$ after this transition}.
On the other hand, for  $|h| \gg 1$, the adiabatic limit is reached instantly and the mode function freezes out immediately whereas the background evolution experiences a transition period before it relaxes to the slow-roll dynamics.

For the sharp transition, the in-in integral in the bispectrum \eqref{bispectinter} can be divided into two nontrivial pieces : one is the contribution from {the instant transition} at $\tau_e$, where $\eta$ can be approximated by the step function as in \eqref{step}; and the other one is the relaxation period from $\tau_e$ to $\tau_0$, which is described by the analytical solution in Section \ref{sec:transition}.

For the first piece, the integral goes from $\tau_{e_-}$ to $\tau_{e_+}$.
At $\tau_{e_-}$, the mode function is described by \eqref{BD}, and $\eta=-6$.
At $\tau_{e_+}$, the mode function is given by \eqref{nonBD}, and $\eta=-6-h$.
Thus taking the squeezed limit and focusing on perturbation modes which exit the Hubble radius during the non-attractor phase,  we can write this contribution to the bispectrum as
\ba \label{bispectrs1}  \small
\lim_{k_3/k \rightarrow 0} B_\calR^a(k, k, k_3)
&=& -{\Im}\calR_{{k}}(\tau_0)^2\calR_{{k}_3}(\tau_0) \int_{\tau_{e_-}}^{\tau_{e_+}}  d\tau{a^2\epsilon\eta'}\left[
\calR_{k}^*(\tau_{e_-})
\calR_{k}^*(\tau_{e_-})
\calR_{k_3}^{*\prime}(\tau_{e_-})\theta(\tau_e-\tau)
\right.\nn\\
&&~~~~\left.+
\calR_{k}^*(\tau_{e_+})
\calR_{k}^*(\tau_{e_+})
\calR_{k_3}^{*\prime}(\tau_{e_+})\theta(\tau-\tau_e)
+{\rm perm.}\right]
\nn\\
&=& \frac{(2\pi)^4}{k_1^3k_3^3} P_\calR^2  \int_{\tau_{e_-}}^{\tau_{e_+}} d\tau \frac{-\eta'}{4}\frac{h\(h+12\)}{(h-6)^2}\[\theta(\tau_e-\tau)+\theta(\tau-\tau_e)\].
\ea
Then via \eqref{step}, the integral above yields
\be
\int_{\tau_{e_-}}^{\tau_{e_+}} d\tau
\frac{h}{4}\frac{h\(h+12\)}{(h-6)^2} \theta'(\tau-\tau_e) \[\theta(\tau_e-\tau)+\theta(\tau-\tau_e)\]
=\frac{h^2}{4} \frac{h+12}{(h-6)^2} ~,
\ee
where in the last step we took an integration by parts to reduce the integral to boundary terms.

The second part of the integral corresponds to the relaxation process after $\tau_e$.
 Substituting the mode function \eqref{nonBD} and \eqref{nonBD2} into \eqref{bispectinter}, its contribution to the squeezed bispectrum is given by
\be \label{bispectrs2}  \small
\lim_{k_3/k \rightarrow 0} B_\calR^b(k, k, k_3)
= \frac{(2\pi)^4}{k_1^3k_3^3} P_\calR^2
\int_{\tau_e}^{\tau_0} d\tau \frac{-\eta'}{8}\sqrt{\frac{\epsilon_0}{\epsilon}}\[2+\eta + \frac{2h}{6-h} \frac{\tau^3}{\tau_e^3} (4+\eta)+ \frac{h^2}{(6-h)^2}\frac{\tau^6}{\tau_e^6} (6+\eta)\]  ~.
\ee
Using the analytical solution during the relaxation \eqref{anasol1} and \eqref{anasol2}, the above integral becomes
\be
\int_{\tau_e}^{\tau_0} d\tau \frac{-\eta'}{8}\sqrt{\frac{\epsilon_V}{\epsilon}}\[2+\eta + \frac{2h}{6-h} \frac{\tau^3}{\tau_e^3} (4+\eta)+ \frac{h^2}{(6-h)^2}\frac{\tau^6}{\tau_e^6} (6+\eta)\]
=-\frac{h}{4} \frac{6h+h^2+12\eta_V}{(6-h)^2} ~,
\ee
where we used $\epsilon_0\simeq\epsilon_V$ which holds in the sharp transition {with $\sqrt{2\epsilon_V}\gtrsim|\eta_V|$.}

Adding these two contributions together, we get
\be \label{ininsharp}
\frac{3}{5}\fnl
= \frac{3 h(h-2\eta_V)}{2 (h-6)^2}  ~.
\ee
As we see, the amplitude of local non-Gaussianity is mainly determined by the $h$ parameter in sharp transition case.
For $|h|\gg1$, it yields the maximum value $\fnl\simeq5/2$, which recovers the result in the initial non-attractor phase.
For the instant transition ($h=-6$), we get a reduced value $\fnl=5/8$.
In general, the sharp transition suppresses the amount of local non-Gaussianity generated during the non-attractor phase.
The extremal case is $h\rightarrow0$, where we have negligible contribution $\frac{3}{5}\fnl = -h\eta_V/12$, similar to the smooth transition result.

Concluding the subsection, we remark that the sharp transition of non-attractor inflation is different from the inflationary feature models, where due to the kink or step on the potential, one may have a short non-slow-roll period which connects two slow-roll stages before and after the local feature.
Since initially inflation is on the slow-roll attractor, long wavelength modes will remain constant during the non-slow-roll period. Therefore these feature models cannot result in nontrivial local non-Gaussianity for large scale perturbations, and the consistency relation is still valid. The reason is that once the mode is frozen in the adiabatic limit it remains so regardless of what may happen after, because a constant is a solution of the EoM for the super-horizon mode function.
However, in the sharp transition here, because of the initial non-attractor phase, long wavelength modes may continue to evolve on super-horizon scales. As a consequence, local non-Gaussianity can be modified on large scales due to the transition.

Related to this issue, it is also known that the presence of sharp feature on potential will generate scale-dependent oscillatory signals in power spectrum and non-Gaussianities (See e.g.~\cite{Chen:2010xka} for a review). The argument is very general and should apply here as well. However, this sinusoidal oscillation starts to appear around the scale $k\sim 1/\tau_e$ and has a wavelength $\Delta k\sim 1/\tau_e$. So they appear at much shorter scales than what we are interested in in this paper.

\subsection{$\delta N$ calculation}
\label{sec:deltaN}

The $\delta N$ formalism \cite{Salopek:1990jq, Sasaki:1995aw, Starobinsky:1986fxa, Sasaki:1998ug, Lyth:2004gb, Lee:2005bb, Lyth:2005fi} is a simple and intuitive approach to the non-linear behaviour of  curvature perturbations.
Based on the separate universe assumption, it mainly captures the super-horizon effects of the perturbation modes, thus it just provides what we need for the calculation of local non-Gaussianity.
For non-attractor inflation, one extra subtlety one should take care of is that the number of e-folds $N$ does not only depend on the initial field value $\phi$, but also on the initial field velocity $\pi$ \cite{Namjoo:2012aa}.
In the following, via $\delta N$ formalism we give a unified calculation of local non-Gaussianity that captures both smooth and sharp transition cases, and recovers the in-in results in Section \ref{sec:smooth} and \ref{sec:sharp} in two extreme limits.

For the non-attractor phase, the number of e-folds $N$ can be easily worked out.
As in Section \ref{sec:transition}, we set $N=0$, $\phi=\phi_e$ and $d\phi/dN=\pi_e$ at the end of the non-attractor phase, then the background equations \eqref{bgEoM} yield the following non-attractor solution in terms of e-folding number $N$
\be
\phi(N) = \phi_e + \frac{\pi_e}{3}\(1-e^{-3N}\)~, ~~~~~~\pi(N)\equiv \frac{d\phi}{dN} = \pi_e e^{-3N}~.
\ee
Next we can invert this solution and obtain the e-folds of the non-attractor phase in terms of the initial $\phi$ and $\pi$
\be
N_i=-\frac{1}{3}\ln \[ \frac{ \pi}{ \pi +3\(\phi-\phi_e\)} \]
=-\frac{1}{3}\ln \frac{\pi}{\pi_e}~,
\ee
where in the second equality we used the following relation of the non-attractor phase
\be \label{relation}
3 \[\phi(N) - \phi_e \] +\pi(N) =\pi_e .
\ee

For the subsequent transition and slow-roll stages, the analytical solutions are already worked out in  \eqref{anasol1} and \eqref{anasol2}. Here we need to study the evolution until the end of the transition, where the slow-roll attractor is  reached. Let us set $N=N_f$ and $\phi=\phi_f$ at that time.
Then $N_f$ is big, and \eqref{anasol1} yields the following approximation
\be
\phi_f  \simeq\frac{s-3-h}{s(s-3)}\pi_e e^{\frac{1}{2}(s-3)N_f} +\frac{2\pi_eh}{s^2-9} +\phi_e~,
\ee
which gives us
\be
N_f \simeq  \frac{2}{s-3} \ln \[\frac{s(s-3)}{s-3-h}\(\frac{\phi_f-\phi_e}{\pi_e} -\frac{2h}{s^2-9} \)\]
 = \frac{2}{s-3} \ln \[\frac{1}{-2\eta_V\pi_e-6\sqrt{2\epsilon_V}} \] + const.
\ee
In the second equality, we separate out the parts unrelated with initial condition $(\phi, \pi)$ as a constant.
Note here, due to the relation \eqref{relation}, $\pi_e$ and also $h$ are determined by the initial $\phi$ and $\pi$ in the non-attractor phase.

Finally, the total e-folding number from the non-attractor phase to the slow-roll stage counted backward in time is given by
\be \label{efolds}
N(\phi, \pi)= N_f -N_i
= \frac{2}{s-3} \ln \[\frac{1}{-2\eta_V\pi_e-6\sqrt{2\epsilon_V}} \] + \frac{1}{3}\ln \frac{ \pi}{\pi_e} + const.
\ee
The $\delta N$ formula is simply given by
\be
\delta N = \frac{\partial N}{\partial \phi} \delta\phi +  \frac{\partial N}{\partial \pi} \delta\pi
+ \frac{1}{2} \frac{\partial^2 N}{\partial \phi^2} \delta\phi^2 +  \frac{\partial^2 N}{\partial\phi\partial \pi}\delta\phi \delta\pi + \frac{1}{2}\frac{\partial^2 N}{\partial \pi^2} \delta\pi^2~.
\ee
Since $\delta\phi$ is approximately constant on super-horizon scales, $\delta\pi$ is exponentially suppressed and thus can be neglected. As a result, from \eqref{efolds} we get
\ba \label{deltaN}
\delta N &=&
\(\frac{\partial N_f}{\partial \phi} - \frac{\partial N_i}{\partial \phi}\) \delta\phi +
\frac{1}{2} \(\frac{\partial^2 N_f}{\partial \phi^2} - \frac{\partial^2 N_i}{\partial \phi^2} \)\delta\phi^2 \\
&=& \(-\frac{1}{\pi_e}+\frac{3}{3\sqrt{2\epsilon_V}+\eta_V\pi_e} \) \delta\phi +  \[ \frac{3}{2\pi_e^2} - \frac{9\eta_V}{2(3\sqrt{2\epsilon_V}+\eta_V\pi_e)^2}\] \delta\phi^2~,
\ea
where again we used the initial condition dependence of $\pi_e(\phi, \pi)$ from \eqref{relation}.
And the local non-Gaussianity directly follows
\be \label{fnldeltaN}
\frac35\fnl=\frac{1}{2}\frac{\partial^2 N}{\partial \phi^2}\left/\(\frac{\partial N}{\partial \phi}\)^2\right.
 = \frac{3 \[4 ({\eta_V}-3) {\eta_V}+h^2+4
   {\eta_V} h\]}{2 (2 {\eta_V}+h-6)^2} 
\ee
This calculation is valid for both smooth transition ($h\to 0$) and sharp transition ($h\neq0$).
As we discussed previously, $\eta_V$ is always small, but $|h|$ can be large for the sharp transition.
Thus similar with the in-in result \eqref{ininsharp}, when $|h|\gg1$, we recover $\fnl=5/2$.
For the smooth transition or sharp transition with small $h$, we get $\fnl\simeq-5\eta_V/6=5\eta_0/12 $, where $\eta_0$ is the second Hubble slow-roll parameter in the slow-roll stage.
{Note that this also agrees with the full in-in calculation. In such cases, the in-in result from cubic interaction term \eqref{interact} is sub-dominant, and thus the leading contribution comes from the field redefinition \eqref{FR}, which yields the same result as above.}

\

We close this section by some concluding remarks. It is interesting to discuss the implications of our results on the consistency relation violation in  canonical non-attractor inflation.
As we know, the power spectrum generated in the non-attractor phase is scale-invariant with $n_s-1=0$.
However, the final result \eqref{fnldeltaN} yields nonzero value for $\fnl$ after the transition.
Even in the smooth transition case where $\fnl$ is slow-roll suppressed, we do not have $\fnl=\frac{5}{12}(1-n_s)$.
Therefore, the consistency relation is still violated in the non-attractor inflation with full consideration of the transition process.

It is also interesting to notice that $\fnl =5/2$ is the maximum non-Gaussianity that one can obtain from such a model irrespective of the details of the transition period. This upper bound holds true even if one considers either a bump potential (where the slope at the transition point is negative) or a step potential (where the potential is discontinuous). That is, although the final $\fnl$ as a function of parameters is clearly different for these cases, its value cannot exceed the $\fnl$ that is generated purely during the non-attractor phase.
{In terms of $\delta N$ formalism, there can be two contributions to the final non-Gaussianity: one from the non-attractor e-folds $N_i$ in \eqref{efolds}, another one from  $N_f$.}
When  $N_i$ terms are the dominant contribution in the $\delta N$ expansion \eqref{deltaN}, we recover the $\mathcal{O}(1)$ non-Gaussianity of the non-attractor phase. In the opposite limit, where $N_f$ terms are dominating, it turns out that the non-Gaussianity is small.
This is an interesting observation without rigorous proof.
But we remark that the $N_f$ part of the evolution is basically the case with non-slow-roll initial condition on a slow-roll potential, which is generically expected to yield small non-Gaussianity, as we argued in Section \ref{sec:general}.
Thus if $N_f$ terms dominate in $\delta N$ expansion \eqref{deltaN}, we expect a slow-roll suppressed $\fnl$.
As a consequence, the upper bound is given by the non-attractor result $\fnl=5/2$ when $N_i$ terms contribute.

Having discussed the difficulties in obtaining larger than $5/2$ non-Gaussianity, it is worth mentioning that it is not impossible in canonical non-attractor models. As a concrete example, consider an upward step in an otherwise constant potential. If the initial velocity of the inflaton is sufficiently high, it climbs up the step during which the non-Gaussianity blows up momentarily but rapidly relaxes to the $\fnl =5/2$ afterwards. Thus, if we cut the potential right after the step and attach it to the slow-roll potential with the condition that the adiabatic limit is reached quickly (i.e. the case with $|h|\gg 1$ as discussed above), we can obtain arbitrarily large non-Gaussianity. Although this is a very restrictive and fine-tuned scenario, it nevertheless shows that there is no physical reason to believe that large local non-Gaussianity cannot be obtained from a canonical, non-attractor, single field model of inflation.
\section{Models with non-canonical kinetic terms}
\label{sec:non-c}

After studying the transition in the canonical ultra-slow-roll inflation, it is also interesting to re-examine the non-canonical model presented in \cite{Chen:2013aj,Chen:2013eea}. We will discuss the background evolution in details. However, since this model cannot be considered as a realistic model of inflation due to the fine tuning of its initial conditions, we study the perturbations only in a specific limit where the analytic calculation is still tractable.

In this model the non-attractor inflation is realized by a k-essence field with the following Lagrangian
\be
\mathcal{L}=P(X, \phi) = X + \frac{X^\alpha}{M^{4\alpha -4}} - V(\phi)~,~~~~~~ V(\phi)=V_0+v\phi^\beta~,
\ee
where $X=-\frac{1}{2}(\partial\phi)^2$, and $\alpha$, $M$, $V_0$, $v$, $\beta$ are free parameters.
In this model, the sound speed $c_s$ is given by
\be \label{cs}
c_s^2 \equiv \dfrac{P_{,X}}{P_{,X}+2 X P_{,XX}} =
\frac{1+\alpha \(\frac{X}{M^4}\)^{\alpha-1}}{1+\alpha (2\alpha-1)\(\frac{X}{M^4}\)^{\alpha-1}} .
\ee
The following variables are also defined here for future reference:
\ba
\Sigma &\equiv& X P_{,X} + 2 X^2 P_{,XX}=\dfrac{X P_{,X}  }{c_s^2}
\, ,\label{sigma}
\\
\lambda &\equiv& X^2 P_{,XX} + \dfrac{2}{3} X^3 P_{,XXX}=\dfrac{X P_{,X}  }{c_s^2} (1-c_s^2)\frac{2\alpha-1}{6} \, .
\label{lambda}
\ea
To the best of our knowledge, so far this is the only model which can give us $\fnl^{\rm local}\gg1$ in single-field inflation with Bunch-Davies initial state.
In this section, we will give a detailed analysis for the transition process in this model, and perform the full calculation to test whether large non-Gaussianity remains or not.

\subsection{Background evolution of k-essence non-attractor model}

First of all, let us focus on the background dynamics of this model.
The equation of motion for inflaton can be written as
\be \label{eom}
\(\frac{\ddot\phi}{c_s^2}+3H\dot\phi\)\[1+\alpha \(\frac{X}{M^4}\)^{\alpha-1}\]+V_\phi=0 ~.
\ee
From the above equation and \eqref{cs} we can see that one important parameter here for the evolution is the ratio $X/M^4$. For $X\gg M^4$, this model is non-canonical with $c_s^2\simeq 1/(2\alpha-1)$; but for $X\ll M^4$, it returns to the canonical case.
In this model initially the inflaton field climbs up the hilltop potential, with the kinetic energy dominated by the non-canonical term. Later on, as $X$ decreases dramatically in the non-attractor phase, the system would go from the non-canonical regime to the canonical regime.

For k-essence field, the slow-roll parameters are expressed as
\ba
\epsilon &\equiv& - \dfrac{\dot H}{H^2} = \dfrac{ X P_{,X}}{ H^2} \, ,\label{epsk}
\\
\eta &\equiv& \dfrac{\dot \epsilon}{H \epsilon} \simeq \dfrac{\ddot \phi}{H
\dot \phi} \left ( 1+ \dfrac{1}{c_s^2}\right)  \, . \label{etak}
\ea
As we know, a non-attractor phase happens when $\epsilon\propto a^{-6}$ and $\eta\simeq-6$.
In the original papers \cite{Chen:2013aj, Chen:2013eea}, an ansatz $\phi(t)\propto a^{\kappa}$ was used to get the initial non-attractor stage.
This was achieved by letting the $V_\phi$ term compete with the $\ddot\phi$ and $\dot\phi$ terms in the equation of motion \eqref{eom}. And the following conditions for parameter choices are required
\be \label{finetune}
\beta = 2 \alpha \, , ~~~~\kappa=\frac{\eta}{2\alpha}~, ~~~~
v =  - \dfrac{M^4}{c_s^2} \left( \dfrac{V_0 \kappa^2}{6 M^4} \right)^\alpha \left(1+\dfrac{3 c_s^2}{\kappa} \right) \, .
\ee

However, it is still not clear how the system transits to the attractor phase in details. In the following we perform a full numerical study of the "non-attractor to slow-roll" transition in the k-essence model. Before that, we summarize the generic behaviour for the evolution first:

\begin{figure}[b]
\centering
\includegraphics[width=0.9\linewidth]{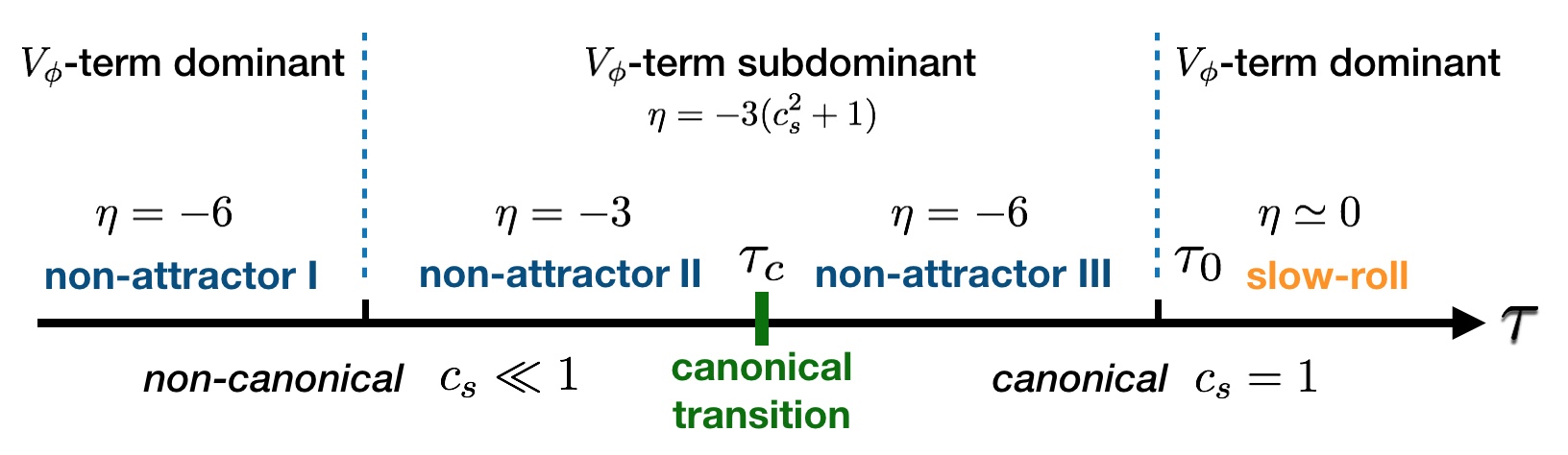}
\caption{The transition process in the k-essence non-attractor model.}
\label{evolve}
\end{figure}

The main results of the numerical solution are shown in Figure~\ref{evolve}.
At the beginning, since the potential is tuned to accommodate with the ansatz as shown in \eqref{finetune}, inflation occurs in the phase with $\eta=-6$, while $X\gg M^4$ gives a small sound speed. We call this initial stage the {\it non-attractor I}.
Then as the inflaton approaches the hilltop, the $V_\phi$ term in \eqref{eom} becomes subdominant, and thus the equation of motion becomes
$\ddot{\phi}+3Hc_s^2\dot\phi\simeq0$, which according to \eqref{etak} yields
\be \label{etacs}
\eta=-3(c_s^2+1)~.
\ee
Since the inflaton field is still  non-canonical ($c_s\ll1$), we have $\eta\simeq-3$. We dub this period as the {\it non-attractor II} phase.
Next, $X$ continues decreasing and becomes smaller than $M^4$, then the canonical term in $P(X, \phi)$ begins to dominate the kinetic energy of inflaton. After that, the scalar field becomes canonical, and we call this moment the {\it canonical transition}. And from \eqref{etacs}, we see the system goes to the canonical non-attractor regime with $\eta=-6$. This stage has the same behaviour with the canonical non-attractor model, and is called  {\it non-attractor III} here.
Finally, the following transition to the slow-roll attractor is the same as what we discussed in Section \ref{sec:canonical}.
As we see, the "non-attractor to slow-roll" transition is much more complicated in the non-canonical model. One important feature is that there is also a canonical transition prior to the slow-roll attractor phase.
This qualitative description is confirmed by the numerical analysis below.

\begin{figure}
\centering
\includegraphics[width=0.6\linewidth]{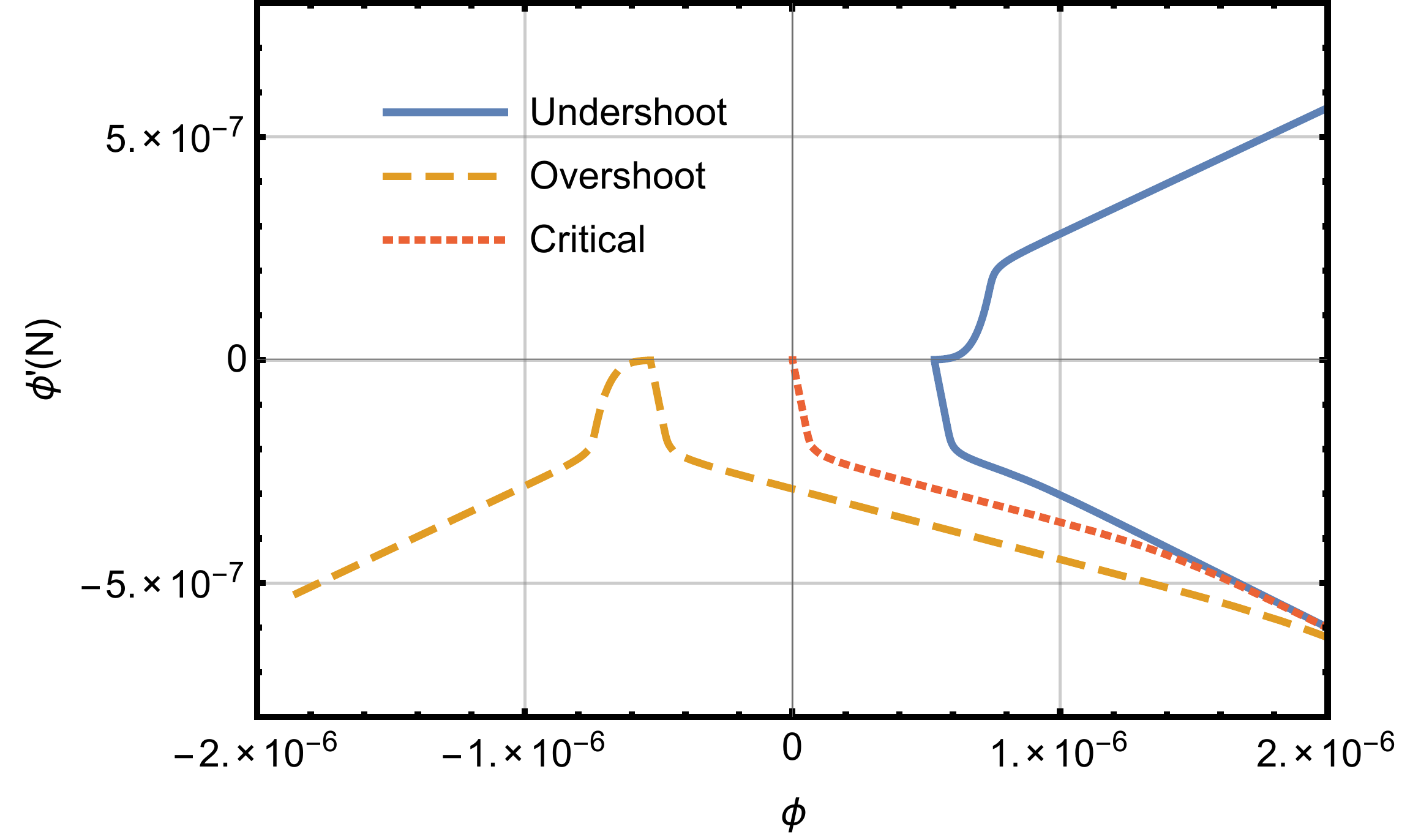}
\caption{The phase space diagram $(\phi, \frac{d\phi}{dN})$ for the k-essence non-attractor model. }
\label{phase}
\end{figure}

\begin{figure}
\centering
\includegraphics[width=0.45\linewidth]{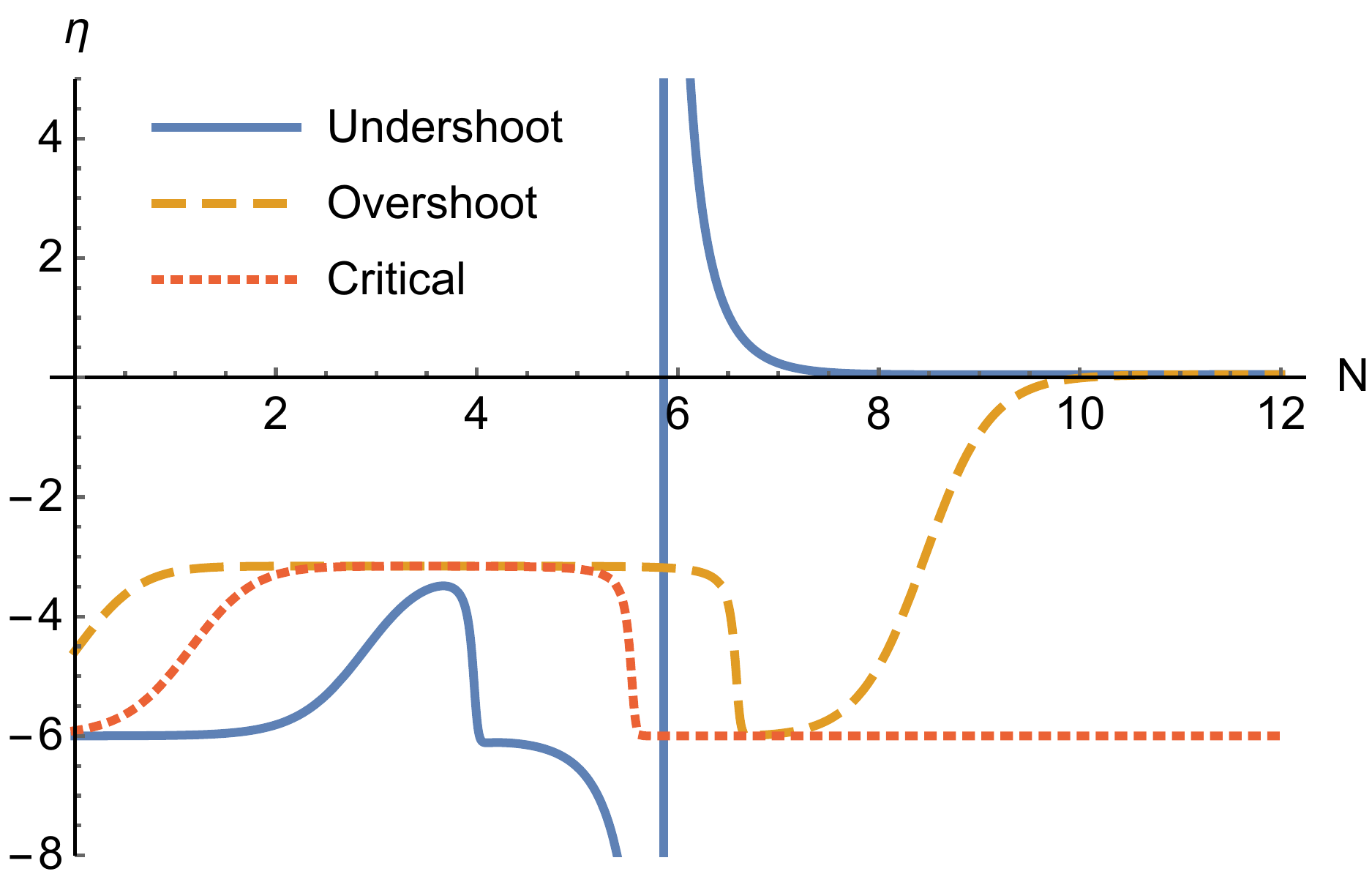} \includegraphics[width=0.45\linewidth]{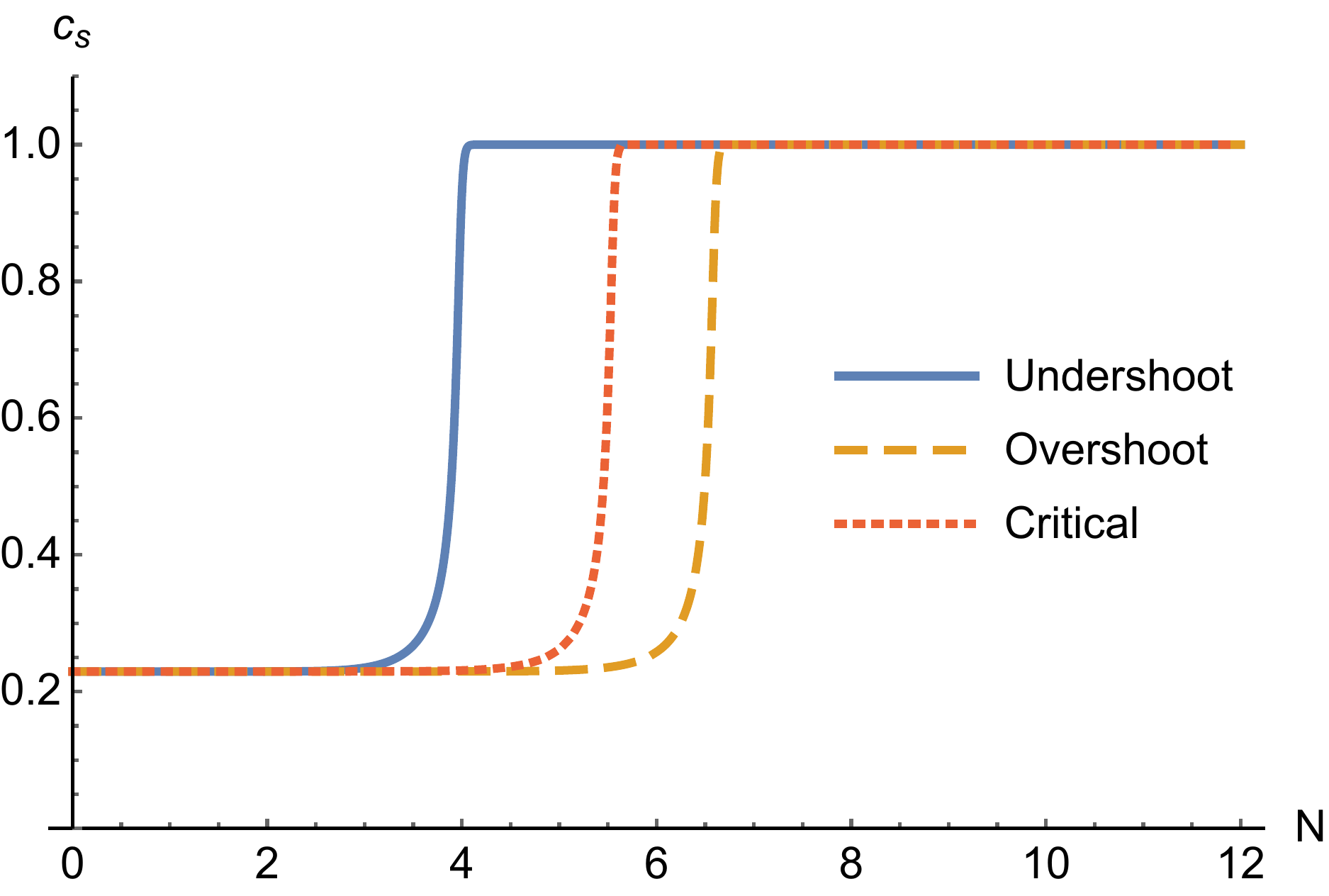}
\caption{The evolution of $\eta$ and $c_s$ in the k-essence non-attractor model. }
\label{cseta}
\end{figure}

\paragraph{Numerical Study.}
Following the choice of parameter values in \cite{Chen:2013aj, Chen:2013eea}, here we take $\alpha=10$, $M=5\times10^{-5}$, $V_0=6.25\times10^{-4}$, while $v$ and $\beta$ are given by the relation in \eqref{finetune}.
Initially inflaton field is set to roll up the hilltop potential from $\phi_i=2\times10^{-6}$. Then via varying the initial field velocity, we find different transition behaviours.
The numerical solutions of background dynamics are shown here.
Figure \ref{phase} gives us the phase space diagram.
In Figure \ref{cseta}, we focus on the evolution of two parameters: the second slow-roll parameter $\eta$, which is important for the non-attractor behaviour, and the sound speed $c_s$, which tells if inflaton field is canonical or not.

From these figures, we can see a generic pattern for the transition process: after the  non-attractor I stage ($\eta\simeq-6$), inflation first enters the non-attractor II phase ($\eta\simeq-3$), and later as shown by the evolution of $c_s$, the canonical transition happens.
Here we introduce a critical field velocity $\dot\phi_c$, for which inflaton can just reach the top of the potential and will stay there forever.
Then accordingly the numerical analysis can be classified into three representative cases:

\begin{itemize}
\item {\it Undershoot} (blue curves). This corresponds to  the case where the initial field velocity is smaller than $|\dot\phi_c|$. After a very short non-attractor II phase, in the canonical non-attractor regime, inflaton stops somewhere before reaching the top of the potential and then rolls backward to initiate the slow-roll phase. At the turning point, since $\dot\phi=0$, we have $\epsilon=0$ and $\eta=\infty$.
\item {\it Critical case} (red curves). The initial field velocity is set to be the critical value. In this case, after the canonical transition, the system reaches an eternal non-attractor stage with $\eta=-6$ and $c_s=1$.
\item {\it Overshoot} (orange curves). This is  the case where the initial field velocity is larger than $|\dot\phi_c|$. As we see, here the non-attractor I stage is very short, while the non-attractor II phase lasts for a longer time, during which the inflaton field rolls over the top of the potential. After this, inflation goes into the canonical non-attractor regime and then transits to the slow-roll stage as we discussed before.
\end{itemize}

In these three cases,  only {\it undershoot} and {\it overshoot} can give us successful "non-attractor to slow-roll" transition.
Although the details can be very different, both these two numerical results verify the evolution in Figure \ref{evolve} and the qualitative description there, i.e. in these non-canonical models the canonical transition always occurs before the relaxation to slow-roll. This holds true at least for our choice of parameters which are consistent with \cite{Chen:2013aj}. It would be interesting to see whether it is also true for other values of the parameters; however, we do not go further in this direction here.
In the following rough calculation of non-Gaussianity, we shall use this general transition behaviour as the basic setup and refer to these two cases (overshoot and undershoot) for details.

\subsection{Non-Gaussianities}

With the above background analysis, we are ready to study the primordial perturbations.
At first glance, a full calculation could be very difficult, since the transition behaviour is quite complicated. Numerical calculation also faces a technical UV-convergence problem because the non-attractor phase is rather short.

However, the problem can be simplified if we focus on the generic pattern of the transition.
As we see, the main difficulty comes from the occurrence of the non-attractor II phase, during which we have $\eta=-3$.
{If this period lasts for a long time (as in the overshoot case), we cannot get a scale-invariant power spectrum for curvature modes that are leaving the horizon during this period.
This can be interesting for the research of features in the primordial perturbations, but in this paper we keep focusing on the behaviour of non-Gaussianity during the transition.}
And for the analysis of the bispectrum, it is the canonical transition that plays a crucial role here.

Therefore we propose the following limit case for analytical study:
The non-attractor II phase is so short such that its effect can be neglected. In this approximation,  before the slow-roll attractor, $\eta$ can always be seen as $-6$ and the canonical transition occurs at some time in this stage. In principle this does not agree with the numerical results since it breaks the relation \eqref{etacs}, but it can be seen as an approximated description of the undershoot case.

Based on the qualitative analysis above, next we focus on the modes which exit the Hubble radius during the non-attractor I stage, and do the back-of-the-envelope estimates for the non-Gaussianities.
The starting point is the cubic action for a general k-essence field in the comoving gauge \cite{Seery:2005wm, Chen:2006nt}.
Since $\epsilon\ll1$ always holds true during the whole transition process, again we can focus on the decoupling limit with only three operators left
\begin{align} \label{action3}
 S_3 \supset \int dt d^3x\[-\frac{a^3\epsilon}{c_s^2}\Xi\frac{\dot\calR^3}{H} -3 \frac{a^3\epsilon}{c_s^4}
 (1-c_s^2)\calR\dot\calR^2 + \frac{a^3\epsilon}{2c_s^2}\frac{d}{dt}
 \bigg(\frac{\eta}{c_s^2}\bigg)\calR^2\dot\calR  \]~.
\end{align}
The coefficient of the $\dot\calR^3$ term is given by
\be
\Xi \equiv 1- \frac{1}{c_s^2} + \frac{2\lambda}{\Sigma}  = \( \frac{2\alpha-1}{3}-\frac{1}{c_s^2}\)(1-c_s^2)~,
\ee
where \eqref{cs}, \eqref{sigma}, \eqref{lambda}, \eqref{epsk} and \eqref{etak} are used for the second equality.
Before the canonical transition we have $\Xi=2(c_s^2-1)/3c_s^2$.
This coefficient and the one for the second term in \eqref{action3} both vanish after the canonical transition.
At the same time, the following field redefinition is also considered in \cite{Chen:2013eea}
\be \label{field_redef}
\calR = \calR_n + \dfrac{\eta}{4 c_s^2} \calR_n^2 + \dfrac{1}{c_s^2 H}
\calR_n \dot \calR_n \, ,
\ee
which can give large non-Gaussianity in the non-attractor I phase. However, since this term should be evaluated in the slow-roll stage where $\eta\simeq0$ and $\dot\calR\simeq 0$, its contribution can be neglected.
For the last operator in \eqref{action3}, again we re-express it via integration by part
\begin{equation}
 -\int dtd^3x \frac{d}{dt}\[\frac{a^3\epsilon}{6c_s^2}\frac{d}{dt}\(\frac{\dot\eta}{c_s^2}\)\] \calR^3 + {\rm surface~ term} ~.
\end{equation}
Plugging in the numerical solution, we confirm that
the effective coupling here is also negligible  during the transitions as in the canonical case.

The difference from the canonical case arises due to a couple of interaction terms in the Lagrangian that are unique for the non-canonical models. Let us estimate their contributions.
The first term in \eqref{action3} gives the bispectrum of
\ba
 B_{\dot \calR^3}(k_1, k_2, k_3) = 12
{\Im} \left[ \calR_{k_1}(\tau_0) \calR_{k_2}(\tau_0) \calR_{k_3}(\tau_0)
\int_{-\infty}^{\tau_c} \frac{a\epsilon d\tau}{c_s^2 H} \, \Xi(\tau) \calR_{k_1}'^*(\tau) \calR_{k_2}'^*(\tau) \calR_{k_3}'^*(\tau)
\right] \, ,
\label{term1}
\ea
where  $\tau_c$ is the conformal time at the canonical transition,
and $\tau_0$ is the one at the beginning of the slow-roll phase.
Since $\Xi$ vanishes after the canonical transition, the in-in integral stops at $\tau_c$.
Another subtlety here is the mode function $\calR_k$.
Since there is a sudden change of $c_s$ around the transition,
in principle one has to use the general slow-roll formalism to solve its behaviour, taking into account the discontinuity around the canonical transition. However, since the integral above vanishes right after the canonical transition, the mode function after transition becomes irrelevant for that integral; and it is easy to check that it does not affect the prefactors $\calR_{k_i}(\tau_0)$ in \eqref{term1} at leading order either. Therefore, for a rough estimate, here we take the following zeroth order approximation
\ba
\calR_k =
\frac{H}{\sqrt{4\epsilon c_s k^3}}(1 + i c_s k \tau)e^{-i c_s k \tau} \, ~~~~
\calR_k' = \frac{H}{\sqrt{4\epsilon c_s k^3}} c_s^2 k^2\tau e^{-i c_s k \tau} -\frac{3}{\tau} \calR_k \,
\label{mode_function}
\ea
Since we mainly care about the modes crossing the Hubble radius during the initial non-attractor phase, we have $-k\tau_0 \ll -k\tau_c \ll 1$. Meanwhile in this limited case we assume $\eta=-6$ before the time $\tau_0$,
which means for this whole period $\epsilon (\tau) =\epsilon_0 \tau^6 /\tau_0^6$.
As a result, the bispectrum becomes
\ba
 B_{\dot \calR^3}(k_1, k_2, k_3) = (2\pi)^4
\(\frac{H^2}{8\pi^2\epsilon_0 c_s}\)^2
\dfrac{3(c_s^2-1)}{4 c_s^2} \(\frac{\tau_0}{\tau_c}\)^6  \frac{k_1^3+k_2^3+k_3^3}{k_1^3 k_2^3 k_3^3}~,
\ea
which is in the local shape.
As we see, when $\tau_0=\tau_c$ it returns to  the previous result in \cite{Chen:2013eea}.
However, if the canonical transition occurs $\Delta N$ e-folds before the slow-roll phase,
$(\tau_0/\tau_c)^6$ would give a suppression factor $\sim e^{-6\Delta N}$.
Correspondingly in the squeezed limit, we get the following amplitude of non-Gaussianity
\ba
 \dfrac{3}{5} \fnl^{\dot \calR^3} = \dfrac{3}{2 c_s^2} (c_s^2-1)\(\frac{\tau_0}{\tau_c}\)^6
  \sim -\dfrac{3}{2c_s^2} (1-c_s^2)  e^{-6\Delta N} \, .
\ea
This suppression is caused by the super-Hubble evolution of the curvature perturbation after the canonical transition.
Since $\calR$ keeps growing until the end of the non-attractor phase, the difference between $\calR(\tau_c)$ and $\calR(\tau_0)$ yields the suppression factor above.

With a similar procedure, the second term in \eqref{action3} gives
\ba
 B_{\calR \dot \calR^2}(k_1, k_2, k_3) = (2\pi)^4
\Big( \frac{H^2}{8\pi^2\epsilon_0 c_s} \Big)^2 \Big(\frac{\tau_0}{\tau_c}\Big)^3 \Big[ 3\(\frac{\tau_0}{\tau_c}\)^3-2 \Big]
\dfrac{3(1-c_s^2)}{8 c_s^2}  \frac{k_1^3+k_2^3+k_3^3}{k_1^3 k_2^3 k_3^3}~,
\ea
Therefore, it is still in the local form and the final amplitude of non-Gaussianity is given by
\ba
 \dfrac{3}{5} \fnl^{\calR \dot \calR^2} = \dfrac{3}{4 c_s^2} (1-c_s^2) \Big(\frac{\tau_0}{\tau_c}\Big)^3 \Big[ 3\(\frac{\tau_0}{\tau_c}\)^3-2 \Big]
  \sim - \dfrac{3}{2c_s^2} (1-c_s^2)  e^{-3\Delta N} \, ,
\ea
where in the last step we ignored the $ e^{-6\Delta N}$ suppression term. Again, the above result agrees with the one in \cite{Chen:2013eea} when $\tau_0=\tau_c$.
In general, the duration of the non-attractor stage after the canonical transition can be $\Delta N \sim \mathcal{O}(1)$, thus the large non-Gaussianity generated in the non-attractor stage can be suppressed a lot.
Summing up the leading terms of two contributions above, we get the following overall amplitude
\be
\frac{3}{5}\fnl \sim   - \dfrac{3}{2c_s^2}  e^{-3\Delta N} ~.
\ee
This estimate shows us how the non-Gaussianity generated in the initial non-attractor stage is suppressed
after the canonical transition.
Notice that the sound speed is determined by the model parameters, while the duration of the non-attractor III stage is related to the choice of initial conditions, thus $c_s$ and $\Delta N$ are two independent parameters. Thus, we conclude that, it is still possible to have large non-Gaussianity in single field inflation.

\section{Conclusion and discussion}
\label{sec:concl}

In this paper, we investigated the production of primordial non-Gaussianities from models of non-attractor inflation.
We revisited various non-attractor models constructed in the literature in order to understand the evolution of large local non-Gaussianity when the models undergo the transition from the non-attractor phase to slow-roll phase. The purpose of this study is less of trying to present these fine-tuned toy-models as phenomenological candidates for data fitting, rather trying to understand more precisely the physical implications of Maldacena's single field consistency relation and various counter-examples that have been constructed.

Comparing with previous studies, we pay special attention to the transition period from the non-attractor phase to the conventional slow-roll phase. Such a transition is necessary for these models to have sufficient efolds or have the correct amplitude of density perturbations.
We considered two types of non-attractor inflation, which are driven by a canonical scalar field and a non-canonical k-essence field, respectively.

For models with canonical kinetic terms, we consider two different evolutionary processes after the non-attractor phase: smooth transition and sharp transition.
Through the calculation of both in-in and $\delta N$ formalism, we find that a full consideration of the transition process generically suppresses the local non-Gaussianity generated in the non-attractor phase, but Maldacena's consistency condition is still violated.
In the smooth transition, the super-horizon modes continue evolving after the non-attractor phase,
and the $\mathcal{O}(1)$ non-Gaussian signals are completely erased during the transition period and the final $\fnl$ at the end of inflation is slow-roll suppressed.
Meanwhile for sharp transition, the final amplitude of the local non-Gaussianity generated in the non-attractor phase depends on the details of the transition process.
In the extremal cases where the curvature perturbation freezes immediately right after the non-attractor phase, we get the maximum possibility of local non-Gaussianity, which recovers the original result in the non-attractor phase $\fnl=5/2$.

For models with non-canonical kinetic terms, although similar situation applies to one of the terms in the Lagrangian, the non-Gaussianities coming from two other terms, which are unique to non-canonical models, survive. Nonetheless, our rough estimations of this case show that the effect of smooth transition is still non-negligible. In addition to the contribution $\sim 1/c_s^2$ obtained in the previous studies, the transition period contributes to an extra suppression factor due to mode evolution outside the horizon during the transition phase. Since these two contributions are independent of each other, the conclusion, that the large local non-Gaussianity can be obtained in such single field models, remains the same; but the expression of $\fnl$ should be revised.

As a final remark, we note that, recently, Ref.~\cite{Bravo:2017gct} argued that the $\mathcal{O}(1)$ local bispectrum generated from the canonical non-attractor inflation model, as calculated in Ref.~\cite{Namjoo:2012aa}, is not locally observable. The study of Ref.~\cite{Bravo:2017gct} focuses on the non-attractor phase. Here we will not analyze their argument in detail which is beyond the scope of this paper. For our purpose, we simply point out that one of the main differences between their work and ours is that we have analyzed in details the subsequent transition process from the non-attractor phase to the standard single field slow-roll inflation, in order to be able to discuss the observability at all. As we have concluded, the final $\fnl$ can range anywhere between zero and a value much larger than 1. If the value of $\fnl$ is much larger than $1-n_s$, these local bispectra should be in principle observable.
At the reheating surface, these local bispectra are indistinguishable from those arising from models in which we replace the single field non-attractor phase with a multifield phase and use the multifield phase to generate the same amount of primordial local bispectra.

\acknowledgments

We are grateful to Ana Ach\'ucarro, Jinn-Ouk Gong, Garret Goon, Enrico Pajer, Gonzalo Palma, Yi Wang and Yvette Welling for helpful discussions.
We would like to thank the authors of Ref.~\cite{Bravo:2017gct} for discussions while both of our works were independently in progress.
YFC is supported in part by the Chinese National Youth Thousand Talents Program, by the NSFC (Nos. 11653002, 11722327, 11421303), by the CAST Young Elite Scientists Sponsorship Program (2016QNRC001), and by the Fundamental Research Funds for the Central Universities.
XC is supported in part by the NSF grant PHY-1417421.
MHN is supported in part by the U.S. Department of Energy under grant Contract Number de-sc0012567.
MS is supported in part by the MEXT KAKENHI Nos. 15H05888 and 15K21733.
DGW is supported by a de Sitter Fellowship of the Netherlands Organization for Scientific Research (NWO).
ZW is supported in part by the Department of Physics at McGill University, and by the Fund for Fostering Talents in Basic Science of the NSFC (No. J1310021).
Part of numerical computations are operated on the computer cluster LINDA in the particle cosmology group at USTC.

\bibliographystyle{JHEP}
\bibliography{bibfile}

\providecommand{\href}[2]{#2}\begingroup\raggedright\begin{thebibliography}{10}

\bibitem{Guth:1980zm}
A.~H. Guth, {\it {The Inflationary Universe: A Possible Solution to the Horizon
  and Flatness Problems}},  {\em Phys. Rev.} {\bf D23} (1981) 347--356.

\bibitem{Linde:1981mu}
A.~D. Linde, {\it {A New Inflationary Universe Scenario: A Possible Solution of
  the Horizon, Flatness, Homogeneity, Isotropy and Primordial Monopole
  Problems}},  {\em Phys. Lett.} {\bf B108} (1982) 389--393.

\bibitem{Starobinsky:1980te}
A.~A. Starobinsky, {\it {A New Type of Isotropic Cosmological Models Without
  Singularity}},  {\em Phys. Lett.} {\bf B91} (1980) 99--102.

\bibitem{Brout:1977ix}
R.~Brout, F.~Englert, and E.~Gunzig, {\it {The Creation of the Universe as a
  Quantum Phenomenon}},  {\em Annals Phys.} {\bf 115} (1978) 78.

\bibitem{Sato:1980yn}
K.~Sato, {\it {First Order Phase Transition of a Vacuum and Expansion of the
  Universe}},  {\em Mon. Not. Roy. Astron. Soc.} {\bf 195} (1981) 467--479.

\bibitem{Fang:1980wi}
L.~Z. Fang, {\it {Entropy Generation in the Early Universe by Dissipative
  Processes Near the Higgs' Phase Transitions}},  {\em Phys. Lett.} {\bf B95}
  (1980) 154--156.

\bibitem{Albrecht:1982wi}
A.~Albrecht and P.~J. Steinhardt, {\it {Cosmology for Grand Unified Theories
  with Radiatively Induced Symmetry Breaking}},  {\em Phys. Rev. Lett.} {\bf
  48} (1982) 1220--1223.

\bibitem{Ade:2015xua}
{\bf Planck} Collaboration, P.~A.~R. Ade et~al., {\it {Planck 2015 results.
  XIII. Cosmological parameters}},  \href{http://arxiv.org/abs/1502.01589}{{\tt
  arXiv:1502.01589}}.

\bibitem{Ade:2015lrj}
{\bf Planck} Collaboration, P.~A.~R. Ade et~al., {\it {Planck 2015 results. XX.
  Constraints on inflation}},  \href{http://arxiv.org/abs/1502.02114}{{\tt
  arXiv:1502.02114}}.

\bibitem{Bartolo:2004if}
N.~Bartolo, E.~Komatsu, S.~Matarrese, and A.~Riotto, {\it {Non-Gaussianity from
  inflation: Theory and observations}},  {\em Phys. Rept.} {\bf 402} (2004)
  103--266, [\href{http://arxiv.org/abs/astro-ph/0406398}{{\tt
  astro-ph/0406398}}].

\bibitem{Liguori:2010hx}
M.~Liguori, E.~Sefusatti, J.~R. Fergusson, and E.~P.~S. Shellard, {\it
  {Primordial non-Gaussianity and Bispectrum Measurements in the Cosmic
  Microwave Background and Large-Scale Structure}},  {\em Adv. Astron.} {\bf
  2010} (2010) 980523, [\href{http://arxiv.org/abs/1001.4707}{{\tt
  arXiv:1001.4707}}].

\bibitem{Chen:2010xka}
X.~Chen, {\it {Primordial Non-Gaussianities from Inflation Models}},  {\em Adv.
  Astron.} {\bf 2010} (2010) 638979,
  [\href{http://arxiv.org/abs/1002.1416}{{\tt arXiv:1002.1416}}].

\bibitem{Wang:2013eqj}
Y.~Wang, {\it {Inflation, Cosmic Perturbations and Non-Gaussianities}},  {\em
  Commun. Theor. Phys.} {\bf 62} (2014) 109--166,
  [\href{http://arxiv.org/abs/1303.1523}{{\tt arXiv:1303.1523}}].

\bibitem{Maldacena:2002vr}
J.~M. Maldacena, {\it {Non-Gaussian features of primordial fluctuations in
  single field inflationary models}},  {\em JHEP} {\bf 05} (2003) 013,
  [\href{http://arxiv.org/abs/astro-ph/0210603}{{\tt astro-ph/0210603}}].

\bibitem{Creminelli:2004yq}
P.~Creminelli and M.~Zaldarriaga, {\it {Single field consistency relation for
  the 3-point function}},  {\em JCAP} {\bf 0410} (2004) 006,
  [\href{http://arxiv.org/abs/astro-ph/0407059}{{\tt astro-ph/0407059}}].

\bibitem{Kinney:2005vj}
W.~H. Kinney, {\it {Horizon crossing and inflation with large eta}},  {\em
  Phys. Rev.} {\bf D72} (2005) 023515,
  [\href{http://arxiv.org/abs/gr-qc/0503017}{{\tt gr-qc/0503017}}].

\bibitem{Namjoo:2012aa}
M.~H. Namjoo, H.~Firouzjahi, and M.~Sasaki, {\it {Violation of non-Gaussianity
  consistency relation in a single field inflationary model}},  {\em Europhys.
  Lett.} {\bf 101} (2013) 39001, [\href{http://arxiv.org/abs/1210.3692}{{\tt
  arXiv:1210.3692}}].

\bibitem{Martin:2012pe}
J.~Martin, H.~Motohashi, and T.~Suyama, {\it {Ultra Slow-Roll Inflation and the
  non-Gaussianity Consistency Relation}},  {\em Phys. Rev.} {\bf D87} (2013),
  no.~2 023514, [\href{http://arxiv.org/abs/1211.0083}{{\tt arXiv:1211.0083}}].

\bibitem{Chen:2013aj}
X.~Chen, H.~Firouzjahi, M.~H. Namjoo, and M.~Sasaki, {\it {A Single Field
  Inflation Model with Large Local Non-Gaussianity}},  {\em Europhys. Lett.}
  {\bf 102} (2013) 59001, [\href{http://arxiv.org/abs/1301.5699}{{\tt
  arXiv:1301.5699}}].

\bibitem{Chen:2013eea}
X.~Chen, H.~Firouzjahi, E.~Komatsu, M.~H. Namjoo, and M.~Sasaki, {\it {In-in
  and $\delta N$ calculations of the bispectrum from non-attractor single-field
  inflation}},  {\em JCAP} {\bf 1312} (2013) 039,
  [\href{http://arxiv.org/abs/1308.5341}{{\tt arXiv:1308.5341}}].

\bibitem{Huang:2013lda}
Q.-G. Huang and Y.~Wang, {\it {Large Local Non-Gaussianity from General
  Single-field Inflation}},  {\em JCAP} {\bf 1306} (2013) 035,
  [\href{http://arxiv.org/abs/1303.4526}{{\tt arXiv:1303.4526}}].

\bibitem{Wands:1998yp}
D.~Wands, {\it {Duality invariance of cosmological perturbation spectra}},
  {\em Phys. Rev.} {\bf D60} (1999) 023507,
  [\href{http://arxiv.org/abs/gr-qc/9809062}{{\tt gr-qc/9809062}}].

\bibitem{Finelli:2001sr}
F.~Finelli and R.~Brandenberger, {\it {On the generation of a scale invariant
  spectrum of adiabatic fluctuations in cosmological models with a contracting
  phase}},  {\em Phys. Rev.} {\bf D65} (2002) 103522,
  [\href{http://arxiv.org/abs/hep-th/0112249}{{\tt hep-th/0112249}}].

\bibitem{Cai:2014bea}
Y.-F. Cai, {\it {Exploring Bouncing Cosmologies with Cosmological Surveys}},
  {\em Sci. China Phys. Mech. Astron.} {\bf 57} (2014) 1414--1430,
  [\href{http://arxiv.org/abs/1405.1369}{{\tt arXiv:1405.1369}}].

\bibitem{Cai:2009fn}
Y.-F. Cai, W.~Xue, R.~Brandenberger, and X.~Zhang, {\it {Non-Gaussianity in a
  Matter Bounce}},  {\em JCAP} {\bf 0905} (2009) 011,
  [\href{http://arxiv.org/abs/0903.0631}{{\tt arXiv:0903.0631}}].

\bibitem{Li:2016xjb}
Y.-B. Li, J.~Quintin, D.-G. Wang, and Y.-F. Cai, {\it {Matter bounce cosmology
  with a generalized single field: non-Gaussianity and an extended no-go
  theorem}},  {\em JCAP} {\bf 1703} (2017), no.~03 031,
  [\href{http://arxiv.org/abs/1612.02036}{{\tt arXiv:1612.02036}}].

\bibitem{Mooij:2015yka}
S.~Mooij and G.~A. Palma, {\it {Consistently violating the non-Gaussian
  consistency relation}},  {\em JCAP} {\bf 1511} (2015), no.~11 025,
  [\href{http://arxiv.org/abs/1502.03458}{{\tt arXiv:1502.03458}}].

\bibitem{Pajer:2017hmb}
E.~Pajer and S.~Jazayeri, {\it {Systematics of Adiabatic Modes: Flat
  Universes}},  \href{http://arxiv.org/abs/1710.02177}{{\tt arXiv:1710.02177}}.

\bibitem{Bravo:2017wyw}
R.~Bravo, S.~Mooij, G.~A. Palma, and B.~Pradenas, {\it {A generalized
  non-Gaussian consistency relation for single field inflation}},
  \href{http://arxiv.org/abs/1711.02680}{{\tt arXiv:1711.02680}}.

\bibitem{Finelli:2017fml}
B.~Finelli, G.~Goon, E.~Pajer, and L.~Santoni, {\it {Soft Theorems For
  Shift-Symmetric Cosmologies}},  \href{http://arxiv.org/abs/1711.03737}{{\tt
  arXiv:1711.03737}}.

\bibitem{Cai:2016ngx}
Y.-F. Cai, J.-O. Gong, D.-G. Wang, and Z.~Wang, {\it {Features from the
  non-attractor beginning of inflation}},  {\em JCAP} {\bf 1610} (2016), no.~10
  017, [\href{http://arxiv.org/abs/1607.07872}{{\tt arXiv:1607.07872}}].

\bibitem{Kallosh:2013hoa}
R.~Kallosh and A.~Linde, {\it {Universality Class in Conformal Inflation}},
  {\em JCAP} {\bf 1307} (2013) 002, [\href{http://arxiv.org/abs/1306.5220}{{\tt
  arXiv:1306.5220}}].

\bibitem{Kallosh:2013yoa}
R.~Kallosh, A.~Linde, and D.~Roest, {\it {Superconformal Inflationary
  $\alpha$-Attractors}},  {\em JHEP} {\bf 11} (2013) 198,
  [\href{http://arxiv.org/abs/1311.0472}{{\tt arXiv:1311.0472}}].

\bibitem{Salopek:1990jq}
D.~S. Salopek and J.~R. Bond, {\it {Nonlinear evolution of long wavelength
  metric fluctuations in inflationary models}},  {\em Phys. Rev.} {\bf D42}
  (1990) 3936--3962.

\bibitem{Sasaki:1995aw}
M.~Sasaki and E.~D. Stewart, {\it {A General analytic formula for the spectral
  index of the density perturbations produced during inflation}},  {\em Prog.
  Theor. Phys.} {\bf 95} (1996) 71--78,
  [\href{http://arxiv.org/abs/astro-ph/9507001}{{\tt astro-ph/9507001}}].

\bibitem{Starobinsky:1986fxa}
A.~A. Starobinsky, {\it {Multicomponent de Sitter (Inflationary) Stages and the
  Generation of Perturbations}},  {\em JETP Lett.} {\bf 42} (1985) 152--155.
  [Pisma Zh. Eksp. Teor. Fiz.42,124(1985)].

\bibitem{Sasaki:1998ug}
M.~Sasaki and T.~Tanaka, {\it {Superhorizon scale dynamics of multiscalar
  inflation}},  {\em Prog. Theor. Phys.} {\bf 99} (1998) 763--782,
  [\href{http://arxiv.org/abs/gr-qc/9801017}{{\tt gr-qc/9801017}}].

\bibitem{Lyth:2004gb}
D.~H. Lyth, K.~A. Malik, and M.~Sasaki, {\it {A General proof of the
  conservation of the curvature perturbation}},  {\em JCAP} {\bf 0505} (2005)
  004, [\href{http://arxiv.org/abs/astro-ph/0411220}{{\tt astro-ph/0411220}}].

\bibitem{Lee:2005bb}
H.-C. Lee, M.~Sasaki, E.~D. Stewart, T.~Tanaka, and S.~Yokoyama, {\it {A New
  delta N formalism for multi-component inflation}},  {\em JCAP} {\bf 0510}
  (2005) 004, [\href{http://arxiv.org/abs/astro-ph/0506262}{{\tt
  astro-ph/0506262}}].

\bibitem{Lyth:2005fi}
D.~H. Lyth and Y.~Rodriguez, {\it {The Inflationary prediction for primordial
  non-Gaussianity}},  {\em Phys. Rev. Lett.} {\bf 95} (2005) 121302,
  [\href{http://arxiv.org/abs/astro-ph/0504045}{{\tt astro-ph/0504045}}].

\bibitem{Seery:2005wm}
D.~Seery and J.~E. Lidsey, {\it {Primordial non-Gaussianities in single field
  inflation}},  {\em JCAP} {\bf 0506} (2005) 003,
  [\href{http://arxiv.org/abs/astro-ph/0503692}{{\tt astro-ph/0503692}}].

\bibitem{Chen:2006nt}
X.~Chen, M.-x. Huang, S.~Kachru, and G.~Shiu, {\it {Observational signatures
  and non-Gaussianities of general single field inflation}},  {\em JCAP} {\bf
  0701} (2007) 002, [\href{http://arxiv.org/abs/hep-th/0605045}{{\tt
  hep-th/0605045}}].

\bibitem{Bravo:2017gct}
R.~Bravo, S.~Mooij, G.~A. Palma, and B.~Pradenas, {\it {Vanishing of local
  non-Gaussianity in canonical single field inflation}},
  \href{http://arxiv.org/abs/1711.05290}{{\tt arXiv:1711.05290}}.

\end{thebibliography}\endgroup

\end{document}